\documentclass{article}
 \usepackage[a4paper,top=2.6cm,bottom=2.6cm,left=2.6cm,right=2.6cm]{geometry}
\usepackage{authblk}
\usepackage{amsmath}
\usepackage{mathtools}
  \usepackage{paralist}
  \usepackage{graphics}
  \usepackage{epsfig} 
\usepackage{graphicx}  
\usepackage{epstopdf}
 \usepackage[colorlinks=true]{hyperref}
\hypersetup{urlcolor=blue, citecolor=red}
\usepackage{csquotes}

\makeatletter
\newcommand{\blx@noerroretextools}{}
\makeatother

\usepackage[sorting=none, style=chem-angew, articletitle=true, maxnames=2]{biblatex}
\usepackage{amsmath}
\usepackage{comment}

\usepackage{xcolor}
\usepackage{accents}
\usepackage{siunitx} 
\usepackage{amsfonts}
\usepackage{amssymb}
\usepackage{booktabs}
\usepackage{amsthm}
\usepackage{mathrsfs}
\usepackage{stackrel,graphicx}
\usepackage{enumitem}
\usepackage{cases}
\usepackage{booktabs}
\usepackage{braket}
\usepackage[strict]{changepage}
\usepackage[small,justification=justified]{caption}
\usepackage{titlesec} 
\usepackage{multicol}
\usepackage{hyperref}
\usepackage{diagbox} 
\usepackage{cleveref}
\usepackage{autonum}
\usepackage{subcaption}

\newtheorem{oss*}{Observation}

\newcommand{\bU}{{\bf U}}

\newcommand{\bmW}{\ensuremath{\vec{\mathcal{W}}}}
\newcommand{\bxi}{\boldsymbol{\xi}}

\newcommand{\bv}{{\boldsymbol{v}}}

\newcommand{\bD}{{\bf D}}
\newcommand{\bV}{{\bf V}}
\renewcommand{\d}{{\, \mathrm d}}
\newcommand{\fperp}{\ensuremath{f^\perp(\tau,\bxi,\hv)}}

\newcommand{\hv}{\hat{\boldsymbol{v}}}

\newcommand{\x}{{\boldsymbol{x}}}

\newcommand{\inner}[2]{\ensuremath{\langle {#1}, {#2}\rangle}}
\newcommand{\Chem}{\mathcal{A}}

\renewcommand{\vec}[1]{\boldsymbol{#1}}
\newenvironment{keywords}
{\noindent\small\textbf{\textit{Keywords---}}}
{}
\usepackage{lineno}
\begin{document}

\title{Hydrodynamic theories of chemotaxis-driven invasion in proliferating cell populations} 
\author[1,2]{Giulia L. Celora}
\author[3]{Martina Conte\thanks{\texttt{Corresponding author: martina.conte@polito.it}}}
\affil[1]{\centerline {\small Wolfson Centre for Mathematical Biology, Mathematical Institute} \newline \centerline{\small  University of Oxford, Oxford OX2 6GG, UK}}
\affil[2]{\centerline {\small School of Mathematics, University of Bristol} \newline \centerline{\small  Fry Building, Bristol BS8 1UG, UK}}
\affil[3]{\centerline{\small Department of Mathematical Sciences ``G.L. Lagrange'', Politecnico di Torino} \newline \centerline{\small  Corso Duca degli Abruzzi, 24 - 10129 Torino, Italy}}

\date{\today}                     
\setcounter{Maxaffil}{0}
\renewcommand\Affilfont{\itshape\small}
\maketitle

\begin{abstract}
Biased migration up chemical gradients and proliferation are fundamental drivers of collective invasion in several biological processes ranging from embryonic morphogenesis to cancer. Nonetheless, our understanding of how their interplay yields distinct invasion patterns remains incomplete. In this work, we propose a multiscale framework to systematically derive macroscopic hydrodynamic theories of cell invasion from a mesoscopic description of cells as biased self-propelled, interacting particles that proliferate. Our framework reveals how clump and stream invasion patterns emerge from the same kinetic equation under different asymptotic regimes of cell proliferation. Stream invasion is characteristic of cell populations in which proliferation balances cell motion. In contrast, clump invasion requires a separation of the hydrodynamic timescale of motion and the slower timescale of proliferation. By means of a multiple-scale approach, our analysis reveals that clump invasion is described as a slow evolution through a family of mass-dependent travelling-wave solutions. Overall, our work offers a novel approach to investigate multiscale regulation of cell invasion in systems where cell proliferation and collective invasion evolve on distinct timescales.
\end{abstract}

\begin{keywords}{Kinetic transport equations, Asymptotic limits, Self-generated chemotaxis, Multiple-scale analysis, Travelling wave fronts}
\end{keywords}

\section*{Introduction}
Collective cell invasion is a fundamental, highly coordinated, and robust process that plays a central role in embryonic development, wound healing, and cancer metastasis~\cite{wu_collective_2025,friedl2009collective,mayor2016front}. In these diverse biological contexts, groups of cells divide, reorganize, and migrate in a coordinated manner into the surrounding environment. These processes are tightly regulated by chemical and mechanical cues that modulate cell polarity, adhesion, motility, as well as cell division~\cite{ridley2003cell}. These feedback mechanisms give rise to robust yet adaptable invasion patterns. Despite being driven by the same underlying biophysical mechanisms, different patterns of invasion can be observed: from stream migration -- such as in neural crest cell invasion~\cite{mclennan2012multiscale,landman2007mathematical,simpson2007cell}, to the cohesive movement of cell clumps or aggregates -- such as in tissues and dense cell collectives~\cite{ford_pattern_2024,celora2026chemotaxiscellaggregatesmorphology,panigrahi_intermittent_2025}. Understanding how the same underlying cellular mechanisms give rise to these distinct invasion modes remains an open challenge.

A wide variety of mathematical models have been developed to understand different aspects of cellular invasion. These approaches span multiple scales, ranging from microscopic (or individual-based) models that resolve migration and proliferation at the single-cell level~\cite{osborne2017comparing,tweedy_self-generated_2016,mclennan2012multiscale,panigrahi_intermittent_2025}, to macroscopic descriptions that capture collective behaviour at the tissue scale through partial differential equations (PDEs) ~\cite{hillen_users_2009,falco_quantifying_2025,ford_pattern_2024,simpson2007cell,ucar_self-generated_2025}. 
Individual-based models explicitly resolve cell-level processes, making them well suited to investigating the biological mechanisms underlying invasion. However, because of their computational complexity, they offer limited analytical insight into the collective invasion dynamics of large populations.
In contrast, continuum PDE models describe tissue-scale behaviour and provide a powerful framework for the analysis of invasion patterns, which often relates to the study of travelling-wave solutions~\cite{simpson_fisherkpp-type_2024,narla_traveling-wave_2021,celora2026chemotaxiscellaggregatesmorphology,celora2026nonlineartheorychemotacticfronts}. However, these models are typically phenomenological, making it difficult to relate their predictions to specific cell-level mechanisms. Hence, the need for a unifying mathematical framework able to rigorously connect these different descriptions and explain how population-level invasion dynamics emerge from the biophysical regulation of cell-level behaviours.

Mesoscopic (or kinetic) models of invasion provide an intermediate description between individual-based and continuum models. At this level, cell motion is commonly represented statistically through a velocity-jump process~\cite{othmer1988models,alt1980biased}, which describes cell movement as a persistent, biased random walk. While these models are often intractable to solve explicitly because of their high-dimensional phase space, different asymptotic methods can be used to derive macroscopic equations under different physical scaling regimes. In particular, in transport-dominated systems, the macroscopic dynamics reduce to a hydrodynamic description in the form of a mass-conservation law for the cell density. This makes the framework effective in rigorously bridging the gap between cellular-level mechanisms and emergent population-level dynamics~\cite{othmer2000diffusion,chalub2004kinetic,eftimie2012hyperbolic}. These mathematical tools have been used to characterise directed migration across various scales. Applications include: long-range chemotactic migration -- \emph{i.e.}, directed cell motion guided by gradients in an external chemoattractant field~\cite{erban2004individual,saragosti_directional_2011,calvez2019chemotactic,chalub2004kinetic}; aggregation via quorum sensing~\cite{painter2002volume}; and invasion into complex anisotropic environments~\cite{chauviere2007modeling,hillen2006m5}. Extensions of the standard kinetic framework have also incorporated cell proliferation~\cite{bellomo2007microscopic,estrada-rodriguez_nonlinear_2025}. 
For mathematical convenience, most upscaling approaches derive macroscopic models under the assumption that proliferation and directed cell motion act on comparable timescales. While this assumption successfully captures stream-like invasion in growing populations~\cite{landman2007mathematical,simpson2007cell,falco_quantifying_2025}, it excludes biologically relevant regimes in which migration is strongly transport-dominated, and proliferation acts only over much longer timescales, such as the cohesive migration of growing cell clumps; specific examples include the zebrafish posterior lateral line primordium~\cite{li2004chemokine} and dense aggregates of social amoebas~\cite{ford_pattern_2024}. 
 
To bridge this gap, we develop a multiscale framework to describe macroscopic invasion of proliferating cells driven by self-generated chemotaxis -- a process where cells create local gradients by consuming an external attractant. Starting from a kinetic description of cells as active, proliferating particles, we systematically derive macroscopic models that capture clump and strand invasion depending on different scaling assumptions for directed cell migration and proliferation. However, standard hydrodynamic asymptotic expansions break down when proliferation evolves on an asymptotically slower timescale than migration. Properly capturing this biologically relevant regime requires the method of multiple scales instead. Our framework therefore reveals how different asymptotic balances between biased cell migration and proliferation give rise to distinct biologically relevant modes of collective cell invasion.

The content of this work is organized as follows. In Section~\ref{sec:kinetic_model}, we introduce our general kinetic model of cell invasion, detailing the functional forms of the operators governing velocity-jump and division mechanisms, and physical cell-cell interactions. In Section~\ref{sec:macroscopic limit}, we analyse the distinct spatio-temporal scales inherent in the system and derive the corresponding macroscopic evolution equations via asymptotic methods.  Finally, in Section~\ref{sec:example}, we provide numerical simulations evaluating two specific regimes: stream migration, which is mediated by the direct balance between proliferation and tactic forces; and clump invasion, which is driven by the balance of active and force drifts and proliferation operates on a much slower timescale. We conclude by discussing our findings and their biological implications.

\section{Kinetic description of active chemotacting particles}\label{sec:kinetic_model}
\begin{figure}[htb]
    \centering
    \begin{subfigure}{0.0\textwidth}
    \captionlistentry{}
    \label{fig:schematicA}
    \end{subfigure}
        \begin{subfigure}{0.0\textwidth}
    \captionlistentry{}
    \label{fig:schematicB}
    \end{subfigure}
        \begin{subfigure}{0.0\textwidth}
    \captionlistentry{}
    \label{fig:schematicC}
    \end{subfigure}
    \includegraphics[width=\linewidth]{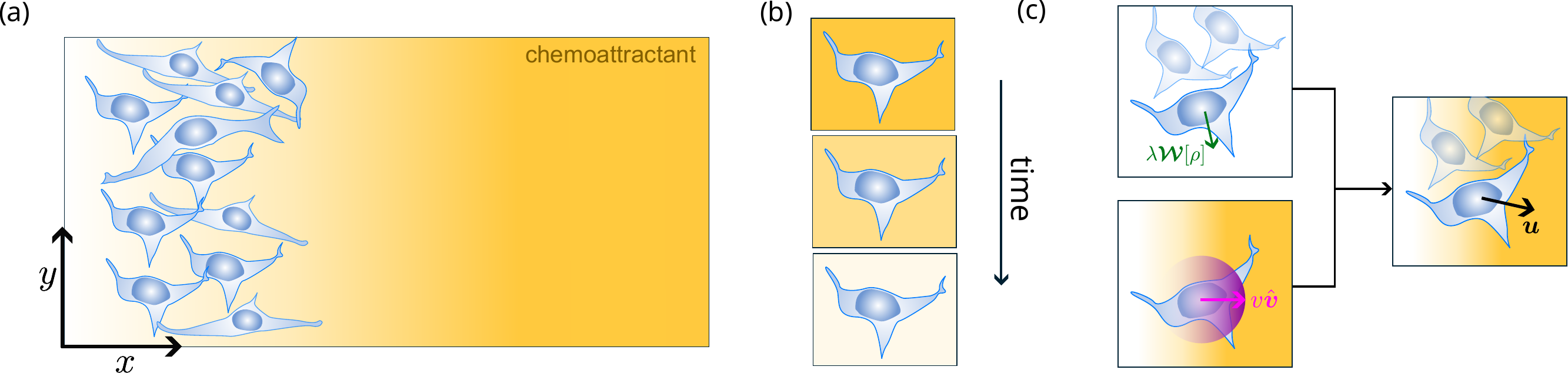}
    \caption{{\bf Schematic of the model setup.} (a) We consider a collection of cells on a 2D substrate that migrate towards the right-hand side of the domain directed by gradients in an external chemoattractant. (b) Cells shape the chemoattractant field via degradation. As a result, they generate the gradients they sense to move. (c) Cell movement is determined by the sum of two effects: (top-right) volume-exclusion $\bmW$ and (bottom-right) self-propulsion along the direction $\hv$, which is sampled from a random probability distribution biased by the chemoattractant.}
    \label{fig:schematic}
\end{figure}

In this work, we develop a minimal multiscale mathematical model to describe long-range collective cell invasion in dilute cell monolayers (\Cref{fig:schematicA}) driven by proliferation and self-generated chemotaxis. This mechanism refers to the biased movement of cells up the gradient of an external chemical signal, which is itself shaped by cellular consumption~\cite{tweedy_self-generated_2016}. 
 
We model cell dynamics at the mesoscopic scale using a kinetic transport equation that accounts for changes in cell orientation in response to a chemoattractant field $\Chem=\Chem(t,\x)$, which cells shape via consumption (Figure~\ref{fig:schematicB}), as well as cell proliferation and volume-filling effects. Specifically, we assume that cells move at a constant propulsion speed $v$ while adjusting their orientation $\hv$ according to a biased velocity-jump process. Since the speed is fixed, these dynamics reduce to jumps in the orientation space, which in the following, we refer to as an {\it orientation-jump} process (Figure~\ref{fig:schematicC}). 
Hence, the cell population is described by the distribution function $f(t,\x,\hv)$, which represents the distribution of cell orientations $\hv\in\mathbb{S}^{n-1}$ for cells at position $\x\in\Omega\subseteq\mathbb{R}^n$ at time $t\ge0$. Here, $\mathbb{S}^{n-1}$ denotes the unit sphere in $\mathbb{R}^n$. The total number of cells at position $\x$ and time $t$ is then described by the macroscopic cell density
\begin{equation}\label{rho_expression}
\rho(t,\x):=\int\limits_{\mathbb{S}^{n-1}} f(t,\x,\hv)\d\hv\,.
\end{equation}
The evolution of the distribution function $f$ is governed by the following kinetic transport equation:
\begin{equation}\label{transport_eq_gen}
	\dfrac{\partial}{\partial t} f(t,\x,\hv) +\nabla_\x\cdot ((v\hv+\lambda \bmW)f(t,\x,\hv)) =\mu\mathcal{L}f(t,\x,\hv) +g\mathcal{P}f(t,\x,\hv)\,.
\end{equation}
The left-hand side of~\eqref{transport_eq_gen} describes cell transport in physical space in a regime of \emph{overdamped dynamics}, where the velocity of an individual, rather than its acceleration, is proportional to the acting forces. This is justified because cell migration occurs in a low-Reynolds-number regime, where viscous friction dominates inertial effects~\cite{carrillo_adhesion_2018}. We define the cell velocity as the superposition of two distinct contributions: self-propelled cell motion and cell-cell physical interactions. The first drift term accounts for the free advective motion of cells with speed $v$ along their polarity direction $\hv$, a standard formulation in kinetic models of persistent random walks. The additional contribution $\lambda\bmW$ captures a mean-field approximation of intercellular (passive) mechanical interactions, \emph{i.e.}, attraction-repulsion dynamics, in the limit where the large-particle numbers. The constant parameter $\lambda\geq 0$ is an effective mobility parameter that captures the effects of the surrounding fluid and extracellular matrix on the cells. The operator $\bmW$ encodes the net direction of the force experienced by a cell and, therefore, the corresponding passive drift velocity, which we consider independent of its active polarisation $\hv$. Similar transport structures arise in kinetic models of active matter and collective biological motion, including models of cell adhesion~\cite{jewell2026cellcelladhesionsustainextended} and quorum sensing~\cite{painter2002volume}.
Following~\cite{carrillo_adhesion_2018,jewell2026cellcelladhesionsustainextended}, we consider the following form for the interaction operator 
\begin{equation}
    \bmW[\rho]=\kappa \nabla(K\star\rho)-\nu\nabla\rho,\label{eq: definition W}
\end{equation}
where the sign $\star$ indicates the convolution of the kernel $K\geq0$ with the density $\rho$. The constants $\kappa \geq 0$ and $\nu >0$ are non-dimensional parameters associated with the strength of long-range attraction and short-range repulsion, respectively. In writing~\eqref{eq: definition W}, we have used the common local approximation for the repulsive interactions, while we allow non-local attractive interactions. 

The first term on the right-hand side of~\eqref{transport_eq_gen} describes the dynamics in the cell orientation space. The parameter $\mu\geq0$ denotes the {\it turning rate}, {\it i.e.,} the rate at which cells change orientation, while the \emph{turning operator} ${\mathcal{L}:L^2(\mathbb{S}^{n-1})\to L^2(\mathbb{S}^{n-1})}$ accounts for the corresponding changes in the distribution $f$ induced by such reorientation events.
Following the standard orientation-jump framework~\cite{hillen2006m5, othmer1988models}, we define $\mathcal{L}$ as a linear integral operator acting on $L^2(\mathbb{S}^{n-1})$ of the form
\begin{equation}\label{turn_operator_gen}
\mathcal{L}f(t,\x,\hv):=T[\Chem](\hv)\rho(t,\x)-f(t,\x,\hv).
\end{equation}
The first term in~\eqref{turn_operator_gen} represents the gain of cells reorienting toward the direction $\hv$, while the second term accounts for the corresponding loss of cells currently oriented along $\hv$ before reorientation. The transition probability $T[\Chem](\hv)$, also referred to as the {\it turning kernel}, describes the probability density that a cell adopts a new orientation $\hv$ after a reorientation event (see shaded purple area in Figure~\ref{fig:schematicC}). For simplicity, we here assume this only depends on the chemoattractant and neglect any cellular memory of the pre-reorientation velocity. As a probability density on $\mathbb{S}^{n-1}$, the turning kernel $T$ must be non-negative and satisfy the normalisation condition \begin{align}
   \int\limits_{\mathbb{S}^{n-1}}T[\Chem](\hv)\d\hv=1,\quad a.e.\quad\x\in\Omega\,,\quad t\ge0\,.\label{condition_T_norm}
 \end{align}
Here, we model the reorientation mechanism through an orientation-jump process whose equilibrium distribution is biased along the gradient of the chemoattractant sensing function $S = S(\Chem)$. The function $S$ encodes the dependence of the reorientation dynamics on the external chemical signal $\Chem$ and its gradients induce a directional bias in the angular dynamics of $\hv$. In line with previous works for local sensing~\cite{ben-ami_using_2024}, we assume that the turning operator takes the exponential form 
\begin{equation}\label{eq:turning operator}
T[\Chem](\hv)= \mathcal{T}\exp\left[-\alpha|\nabla S(\Chem)|\left(1-\frac{\inner{\hv}{\nabla S(\Chem)}}{|\nabla S(\Chem)|}\right)\right],
\end{equation}
where $\mathcal{T}$ is a normalisation constant ensuring~\eqref{condition_T_norm}, and $\alpha > 0$ is the parameter controlling the sharpness of the distribution, with variance scaling as $\mathcal{O}(\alpha^{-2})$.  We notice that, in the limit of shallow gradients $\alpha|\nabla S|\ll1$, a first-order expansion of~\eqref{eq:turning operator} yields a linear dependence on the signal gradient, recovering the classical weakly biased turning kernel introduced in~\cite{othmer2002diffusion}, in which the directional bias appears as a perturbation of the random diffusive motion (see~\Cref{App:DiffLimit} for detailed discussion of this regime).

Finally, the operator $\mathcal{P}f$ on the right-hand side of~\eqref{transport_eq_gen} incorporates changes in the distribution function due to cell proliferation, which occurs at a characteristic rate $g \geq 0$. We assume that proliferation is independent of cell orientation and that daughter cells inherit the polarisation of the mother cell. Under these assumptions, the action of $\mathcal{P}$ reduces to a scaling of the local distribution, namely 
\begin{equation}\label{eq:proliferation kernel}
\mathcal{P} f(t,\x,\hv) = G[\rho, \Chem] f(t,\x,\hv)\,,
\end{equation}
where the function $G$ captures how the cell proliferation rate is affected by the local cell density via cell crowding and/or the external chemical signal $\Chem$. Since we assume that cell number increases only due to proliferation, we couple~\eqref{transport_eq_gen} with mass-conservative boundary conditions by requiring the total normal flux across the boundary $\partial\Omega$ to vanish, namely
\begin{equation}\label{eq: general no flux}
\int_{\mathbb{S}^{n-1}}(v\hv+\lambda\bmW)f(t,\x,\hv)\cdot \boldsymbol{\hat n}(\x)d\hv=0,\qquad\forall\x\in \partial\Omega\,,\,\,\,t>0
\end{equation}
where $\boldsymbol{\hat n}$ denotes the outward unit normal vector on $\partial\Omega$. This ensures conservation of the total cell mass in the absence of proliferation.

For the chemoattractant field $\Chem$, we assume that gradients are generated by the cells via consumption and describe the time evolution of the chemoattractant field  $\Chem$ via the following reaction-diffusion equation
\begin{equation}
    \frac{\partial \Chem}{\partial t}=D_\Chem\nabla^2\Chem-\bar{\beta} \beta_\Chem[\rho,\Chem]\Chem.\label{eq:chemoattractant}
\end{equation}
In Eq.~\eqref{eq:chemoattractant}, $D_{\Chem}>0$ is the diffusion coefficient of the chemical, $\bar{\beta}>0$ is the characteristic per-cell chemoattractant consumption rate, and the function $\beta_\Chem$ captures how consumption depends on the local cell density $\rho$ and the chemoattractant concentration itself $\Chem$. 

A macroscopic description of the cell population can be formally derived from the kinetic framework by taking suitable moments of the distribution function~\cite{othmer1988models,hillen2006m5}. In particular, integrating~\eqref{transport_eq_gen} over the orientation space $\mathbb{S}^{n-1}$ and using the normalisation condition~\eqref{condition_T_norm}, we obtain the macroscopic continuity equation
\begin{equation}\label{eq:macroscopic evolution density}
    \frac{\partial\rho}{\partial t}+\nabla_{\x}\cdot\left(\rho\bar{\vec{v}}_c+\lambda\bmW[\rho]\rho\right)=gG[\rho, \Chem]\rho,
\end{equation}
where the macroscopic active contribution to the cell flux is defined as the first moment of the distribution function,
\begin{equation}\label{eq:macroscopic cell flux}
\rho \bar{\vec{v}}_c=\int\limits_{\mathbb{S}^{n-1}}v\hv\,f(t,\x,\hv)\d\hv\,.
\end{equation}
Since the turning kernel depends on the chemoattractant field $\Chem$, the resulting flux implicitly encodes chemotactic effects via an anisotropy in the velocity distribution. Eq.~\eqref{eq:macroscopic evolution density} is not closed at the level of the first moment, since the flux $\rho\bar{\vec{v}}_c$ requires the full distribution $f$. A corresponding evolution equation for momentum $\rho\bar{\vec{v}}_c$ can be obtained by multiplying~\eqref{transport_eq_gen} by $\hv$ and integrating over $\mathbb{S}^{n-1}$. However, the resulting equation depends on higher-order moments of $f$. To address this problem, several methodologies have been proposed in the literature, including classical moment closure techniques and asymptotic scaling expansions~\cite{hillen2013transport}. In the following section, we exploit asymptotic techniques to derive a closed-form evolution equation for the cell density~\eqref{eq:macroscopic evolution density} under different assumptions on the relative scaling between the biological timescales involved in the model. 

\section{Formal derivation of the macroscopic limits}\label{sec:macroscopic limit}
To elucidate the macroscopic mechanisms driving cell invasion, we derive a closed-form evolution equation for the cell density $\rho$~\eqref{eq:macroscopic evolution density} under different asymptotic regimes. The derivation of macroscopic limits for orientation-jump processes is a well-established field in mathematical biology, typically involving the transition from kinetic descriptions to either parabolic or hyperbolic continuum models~\cite{othmer2000diffusion, othmer2002diffusion, chalub2004kinetic, eftimie2012hyperbolic}. Importantly, the form of the resulting macroscopic equations depends on the relative scaling between transport, turning, and interaction terms, which determines the appropriate closure regime. 

Starting from the mesoscopic framework, we identify the relevant asymptotic regimes through a systematic nondimensionalisation of the governing equations. Specifically, to identify the physical balances governing the population dynamics over macroscopic spatial and temporal scales, we  define the following dimensionless space and time variables:
\begin{equation}\label{gen_scal}
\quad \bxi=\dfrac{\x}{\ell}\,,\quad \tau=\dfrac{t}{\tilde{t}},
\end{equation}
where $\ell$ and $\tilde{t}$ indicate the characteristic macroscopic length and time scale, respectively. Substituting these variables into the kinetic equation~\eqref{transport_eq_gen} and the chemoattractant dynamics~\eqref{eq:chemoattractant}, we obtain the rescaled equations:
\begin{align}\label{eq:f rescaled}
        \text{St} \frac{\partial f}{\partial \tau}+ \nabla_{\bxi} \cdot((\hv+\Lambda\bmW) f)&=\frac{1}{\text{Kn}}\mathcal{L}f+ 
        \text{St}\Phi\mathcal{P}f\,,\\[0.2cm]
         \dfrac{1}{\tau_{\Chem}}\dfrac{\partial \Chem}{\partial \tau}&=\nabla_{\bxi}^2\,\Chem-\text{Da}\beta_\Chem[\rho,\Chem]\Chem\,.\label{macroscopic chemo}
    \end{align}%
From this scaling procedure, we identify the following key dimensionless parameters governing the evolution of our system. For the cell dynamics, we obtain:
\begin{enumerate}
\item[(i)] the {\it Knudsen number} $\text{Kn}:=v/(\ell\mu)$, which characterises the ratio between the mean free path of a cell $\ell_v=v/\mu$ (the average distance travelled between two reorientation events), and the macroscopic length scale of the system $\ell$; 
\item[(ii)] the \emph{Strouhal number} $\text{St}:=\ell/{v}\tilde{t}$, which quantifies the ratio between the advective timescale $ t_{v}=\ell/{v}$ and the macroscopic observation time $\tilde{t}$; 
\item[(iii)] the dimensionless proliferation rate $\Phi:=g\tilde{t}$, which is the ratio between the macroscopic observation time $\tilde{t}$ and the characteristic timescale of cell division $ t_p=g^{-1}$; 
\item[(iv)] the relative interaction strength $\Lambda:=\lambda/{\ell v}$, which is the ratio between the speeds associated with attraction-repulsion-induced drift $v_{\bmW}=\lambda/\ell$ and active self-propulsion $v$.
\end{enumerate}
Conversely, for the chemoattractant dynamics, we have:
\begin{enumerate}
\item[(v)] the \emph{Damk\"{o}hler number} $\text{Da}:=\bar{\beta}\ell^2/D_\mathcal{A}$, which measures the ratio between chemoattractant consumption and diffusion;
\item[(vi)] the dimensionless chemoattractant relaxation timescale $\tau_{\Chem}:=\tilde{t}D_{\Chem}/\ell^2$, which quantifies the ratio between the macroscopic time scale $\tilde{t}$ and the timescale of chemoattractant diffusion  $t_{\Chem}=\ell^2/D_\Chem$.
\end{enumerate}
Here, we set the macroscopic length scale $\ell$ to coincide with the characteristic chemoattractant decay length $\ell_\Chem=:\sqrt{D_\Chem/\overline{\beta}}$, yielding Da$=1$. Following experimental evidence for eukaryotic cells and bacteria~\cite{li2008persistent,allen2020cell}, we find that the mean free path ${\ell_v \approx 10-30\, \mu\text{m}}$ is significantly smaller than the characteristic length scale of a chemoattractant, $\ell_\Chem \approx 10^2-10^3 \mu\text{m}$~\cite{hofer1995dictyostelium,ibanes2008theoretical}. This separation of scales justifies the asymptotic regime $\text{Kn} = \varepsilon \ll 1$. This regime corresponds to a scenario where the turning frequency $\mu$ is sufficiently high such that the distribution of orientations relaxes rapidly toward a local equilibrium.
Regarding the relative interaction strength, we assume $\Lambda \sim \mathcal{O}(1)$, implying that attraction–repulsion dynamics operate on the same timescale as active self-propulsion. Physically, this means that in dense environments, intercellular forces—such as adhesion and repulsion—are strong enough to modulate individual trajectories at the leading order.

The asymptotic behaviour of \eqref{eq:f rescaled} in the $\text{Kn} \to 0$ limit depends on the relative scaling between the Strouhal and Knudsen numbers, which is set by the choice of the macroscopic time scale $\tilde{t}$. 
In the following, we focus on the \textit{hydrodynamic (hyperbolic) regime}, which is particularly well suited for describing directed collective motion and the propagation of coherent cellular fronts. In this limit, the macroscopic time scale is assumed to be comparable to the characteristic time of advective transport, {\it i.e.,}
\begin{equation}\label{eq:hyperbolic scaling}
    \tilde{t} \sim t_v = \frac{\ell}{v}.
\end{equation}
Considering the characteristic macroscopic length scale $\ell \sim 10^2-10^3 \mu\text{m}$ and a typical single-cell migration speed $v \sim 1\text{--}5\,\mu\text{m/min}$~\cite{mclennan2012multiscale,ford_pattern_2024}, this advective time scale yields an estimate of $\tilde{t} \in 10-200$ minutes. By recalling the definition of the Knudsen number, {\it i.e.,} $\text{Kn} = v/(\mu \ell) = \varepsilon$, and given~\eqref{eq:hyperbolic scaling}, the Strouhal number is an $\mathcal{O}(1)$-quantity in the $\text{Kn}\to 0$ limit. Furthermore, given that the advective transport is driven by a self-generated chemotactic mechanism, the hyperbolic regime~\eqref{eq:hyperbolic scaling} naturally yields the scaling $\tau_{\Chem} = \mathcal{O}(1)$.

Concerning proliferation, we expect the magnitude of $\Phi$ to vary between different biological systems. When $\Phi = \mathcal{O}(1)$, proliferation and macroscopic drift operate on the same timescale. In this case, the growth term $\Phi\, G[\rho,\Chem] \rho$ enters the macroscopic equation as a leading-order contribution. Biologically, this reflects the stream invasion in development~\cite{mclennan2012multiscale}, where growth directly drives the advancement of the front. In contrast, when $\Phi= \mathcal{O}(\varepsilon)$, cell movement is significantly faster than the proliferative rate. This regime characterises the cohesive migration of cell clumps or aggregates~\cite{ford_pattern_2024,panigrahi_intermittent_2025}, where rapid, chemotaxis-driven displacement dominates the short-term dynamics. Since we are interested in capturing both these modes of invasion, we define
\begin{equation}
\Phi=\Phi(\varepsilon):=\phi_0+\varepsilon\phi_1,\label{eq:definition Phi}
\end{equation}
where $\phi_{0,1}$ are non-negative constants. Then, the first regime corresponds to $\phi_0>0$, $\phi_1=0$, while the second regime to $\phi_1>0$ and $\phi_0=0$.

\subsection{A hydrodynamic theory of invasion}\label{sec:hyperbolic scaling standard}
To resolve the initial fast dynamics, we analyse the behaviour of the rescaled transport equation~\eqref{eq:f rescaled} under the hyperbolic scaling~\eqref{eq:hyperbolic scaling}; namely,
\begin{equation}\label{transport_eq_rescale_hyp}
 \varepsilon \frac{\partial f}{\partial \tau}(\tau,\bxi,\hv)+ \varepsilon\nabla_{\bxi} \cdot( (\hv+\Lambda \bmW) f(\tau,\bxi,\hv))=\mathcal{L}f (\tau,\bxi,\hv)+ 
        \varepsilon\Phi(\varepsilon)\mathcal{P}f(\tau,\bxi,\hv)\,.
\end{equation}
Following the classical framework for deriving macroscopic hydrodynamic descriptions from the underlying kinetic models~\cite{hillen2013transport,dolak2005kinetic,ben-ami_using_2024}, we adopt a Chapman–Enskog expansion of the distribution function $f$:
\begin{equation}\label{eq:chapman expansion}
f(\tau,\bxi,\hv)=\rho(\tau,\bxi)T[\Chem](\hv) +\varepsilon \fperp \,.
\end{equation}
Here, the leading-order term of the expansion makes the turning operator~\eqref{turn_operator_gen} vanish. All the functions nullifying the turning operators are proportional to $T$ up to a multiplicative constant, which is independent of the orientation $\hv$. As a result, the subspace of $L^2(\mathbb{S}^{n-1})$ defined by $\langle T(\hv)\rangle$ is the kernel of the turning operator $\mathcal{L}f$, and it has dimension equal to one. To ensure the uniqueness of the decomposition of $f$, we impose the orthogonality condition $f^{\perp}\in\langle T(\hv)\rangle^{\perp}$, which implies that the correction term carries no mass \cite{hillen2013transport}. Substituting the expansion~\eqref{eq:chapman expansion} into \eqref{transport_eq_rescale_hyp} and integrating over the orientation space $\mathbb{S}^{n-1}$, we find that the evolution of the macroscopic density $\rho$ is governed by
\begin{equation}\label{eq:density_advection_chapman_expansion}
    \frac{\partial \rho}{\partial \tau} +\nabla_{\bxi} \cdot \Big[\rho\Big(\bV_T[\Chem]+\Lambda \bmW[\rho]\Big)\Big]+\varepsilon \nabla_{\bxi} \cdot\int\limits_{\mathbb{S}^{n-1}} \hv \fperp d\hv=\Phi\, G[\rho, \Chem] \rho,
\end{equation}
where the macroscopic advection velocity $\bV_T[\Chem]$ is the first moment of the turning kernel
\begin{equation}\label{UT}
\bV_T[\Chem]:=\int\limits_{\mathbb{S}^{n-1}}\hv T[\Chem](\hv)\d\hv\,.
\end{equation}
The term $\bV_T$ captures how cellular re-polarisation affects the macroscopic cell flux. As detailed in \Cref{app:fperp_derivation}, in the small $\varepsilon$-limit, the first-order correction to the cell flux~\eqref{eq:density_advection_chapman_expansion} can be approximated as:
\begin{equation}\label{eq:flux fperp}
    -\int\limits_{\mathbb{S}^{n-1}} \hv \fperp d\hv\sim\nabla_{\bxi} \cdot\left(\bD_T\rho \right) +\rho \left(\dfrac{\partial \bV_T}{\partial \tau}+\left(\bV_T+\Lambda \bmW[\rho]\right)\nabla_{\bxi}\cdot \bV_T\right),
\end{equation}
where $\bD_T=\bD_T[\Chem]$ is the variance-covariance matrix of the turning kernel
\begin{align}
\bD_T[\Chem]:=\int\limits_{\mathbb{S}^{n-1}}(\hv-\bV_T[\Chem])\otimes (\hv-\bV_T\Chem])T[\Chem](\hv)d\hv\,.\label{VT}
\end{align}
Thus, plugging~\eqref{eq:flux fperp} into~\eqref{eq:density_advection_chapman_expansion} and collecting the leading- and first-order contributions, we obtain the following macroscopic continuity equation for the dynamics of the cell density $\rho$, accurate up to order $\mathcal{O}(\varepsilon)$:
%\begin{subequations}\label{macro_mass_chemo_hydrodynamics}
\begin{equation}
    \frac{\partial \rho}{\partial \tau}- \nabla_{\bxi}\cdot\left[ \varepsilon\nabla_{\bxi}\cdot\left(\bD_T[\Chem]\rho\right)  -\left(\bU[\rho,\Chem]+\varepsilon \tilde{\bV}[\rho,\Chem]\right)\rho\right]-\Phi(\varepsilon) G[\rho,\Chem]\rho\,=%\sim 
    0\,.\label{macroscopic mass balance hydrodynamics}
\end{equation}
Here 
\begin{align}
    \bU[\rho,\Chem]&:=\bV_T[\Chem]+\Lambda \bmW[\rho],\label{eq: definition total cell velocity}\\
    \tilde{\bV}[\rho,\Chem]&:=-\bU[\rho,\Chem] \nabla_{\bxi}\cdot \bV_T[\Chem]-\frac{\partial \bV_T[\Chem]}{\partial \tau}\,,\label{eq: definition V tilde}
\end{align}
and $\mathcal{A}$ satisfies
\begin{equation}
     \dfrac{1}{\tau_{\Chem}}\dfrac{\partial \Chem}{\partial \tau}=\nabla_{\bxi}^2\,\Chem-\beta_\Chem[\rho,\Chem]\Chem\,.\label{macroscopic chemo rescale}%
\end{equation}
Eq.~\eqref{macroscopic mass balance hydrodynamics} is subject to the boundary conditions
\begin{equation}\label{eq: general no flux_macro}
\int_{\partial\Omega}\rho\bU[\rho,\Chem]\cdot \boldsymbol{\hat n}\d s+\varepsilon\int_{\partial\Omega}\rho\left(\nabla_{\bxi}\cdot\left(\bD_T[\Chem]\rho\right)- \tilde{\bV}[\rho,\Chem]\right)\cdot \boldsymbol{\hat n}\d s=0,
\end{equation}
where $\boldsymbol{\hat n}$ denotes the normal outward unit vector on $\partial\Omega$. Here, $\bU$ corresponds to the leading-order contribution to the advection velocity, which incorporates the leading-order drift $\bV_T$~\eqref{UT} related to the orientation-jump process and the contribution of the intercellular forces $\bmW$~\eqref{eq: definition W}. Specifically, $\bV_T$ can be interpreted as the average polarisation field at a given location $\bxi$ and time $\tau$. Instead, the diffusion tensor $\bD_T$ captures the stochastic nature of the cell re-polarisation process, while the corrections to the drift velocity $\tilde{\bV}$ account for the role that spatio-temporal modulations of $\bV_T$ play in shaping the macroscopic cell flux. Hence, the hyperbolic scaling~\eqref{eq:hyperbolic scaling} ensures that the macroscopic cell flux enters the conservation law at the leading order, capturing the persistent nature of cellular trajectories before the long-term diffusive effects due to the stochastic nature of the polarisation process cease to be asymptotically negligible. 
%\end{subequations} 

Since we are interested in models of long-range migration, we focus on the scenario in which~\eqref{macroscopic mass balance hydrodynamics}-\eqref{macroscopic chemo rescale} admits a travelling-wave solution in the $\varepsilon=0$ regime:
\begin{equation}\label{eq:travelling wave ansatz}
\lim_{\tau\to\infty} \varphi_0(\tau,\bxi)=\bar{\varphi}_0(\bxi-\vec{c}_0\tau), \quad \varphi=\rho,\Chem,
\end{equation}
where $\vec{c}_0$ is the velocity of the travelling wave, and the subscript $0$ indicates the solution to~\eqref{macroscopic mass balance hydrodynamics}-\eqref{macroscopic chemo rescale} when $\varepsilon =0$. This behaviour is commonly observed in systems where self-generated chemotaxis is coupled with cell proliferation~\cite{narla_traveling-wave_2021,cremer_chemotaxis_2019}. The dynamics of cell invasion is further influenced by the parameter $\Phi$~\eqref{eq:definition Phi}, which controls the role of cell proliferation on the hyperbolic scale. Dealing with the regimes $\phi_0>0$ and $\phi_0=0$ requires different asymptotic techniques. In particular, as explored in Section~\ref{sec:multiple scale}, the interplay between fast movement and slower proliferation (\emph{i.e.}, $\phi_0=0$ in~\eqref{eq:definition Phi}) has to be handled via the method of multiple scales. By introducing distinct fast and slow temporal variables, we can derive the appropriate governing equations for the corresponding macroscopic invasion dynamics.

\subsection{Leading-order dynamics: drift-force-proliferation balance}\label{sec:stand_drift_prol}
Under the assumption of a balance between drift, interaction forces, and proliferation, {\it i.e.,} when ${\Phi=\mathcal{O}(1)}$, the macroscopic dynamics are governed by the direct competition between convective flux and population growth. As discussed in~\Cref{sec: result proliferation-driven invasion}, these ingredients are sufficient to sustain cellular invasion. In this scenario, on the hyperbolic time scale, the advective and reaction terms dominate over the spatial dispersion. Therefore, we consider a standard asymptotic expansion in powers of $\varepsilon$ of the 
solutions to \eqref{macroscopic mass balance hydrodynamics}, namely
\begin{equation}\label{eq:general_asymptotic_expansion_eps}
\varphi(\tau,\bxi)=\sum_{n=0}^\infty \varepsilon^n\varphi_n(\tau,\bxi)\,,\quad \varphi\in\left\{\rho,\Chem\right\}\,,
\end{equation}
%\begin{equation}\label{eq: asymptotic Hilbert expansion}
%\varphi=\varphi_0+\varepsilon\varphi_1+\mathcal{O}(\varepsilon^2) \quad \varphi=\{\rho,\Chem\}.
%\end{equation}
Then, the leading-order spatio-temporal dynamics of the cell density $\rho_0$ is well captured by taking the formal hyperbolic limit $\varepsilon \to 0$ in Eq. \eqref{macroscopic mass balance hydrodynamics}, yielding:
\begin{equation}
    \frac{\partial \rho_0}{\partial \tau}+\nabla_{\bxi}\cdot\left[\bU[\rho_0,\Chem_0]\rho_0\right]=\phi_0\, G[\rho_0,\Chem_0]\,\rho_0,\label{eqref: hydro leading order with prol}
\end{equation}
subject to no-flux boundary conditions 
\begin{equation}\label{eq: no flux_macro leading order}
    \int_{\partial\Omega}\rho_0 \bU[\rho_0,\Chem_0]\cdot \hat{\vec{n}} \d s=0,
\end{equation}
where $\bU$ is as defined in~\eqref{eq: definition total cell velocity}.
The evolution for the leading-order behaviour of the chemoattractant concentration is similarly obtained from~\eqref{macroscopic chemo}
\begin{equation}\label{eqref:hydro leading order with prol A}
    \dfrac{1}{\tau_{\Chem}}\dfrac{\partial \Chem_0}{\partial \tau}=\nabla_{\bxi}^2\,\Chem_0-\beta_\Chem[\rho_0,\Chem_0]\Chem_0\,.
\end{equation}%
Eq.~\eqref{eqref: hydro leading order with prol} represents a pure drift-reaction dynamics, where proliferation directly guides the spatial invasion process. The $\varepsilon$-order corrections derived in Eq.~\eqref{macroscopic mass balance hydrodynamics} are typically negligible compared to the dominant drift and reaction terms. However, these higher-order contributions may become physically relevant in regions where 
\begin{equation}\label{diff_condition}
\nabla_{\bxi}\cdot\left[\bU[\rho_0,\Chem_0]\rho_0\right]-\phi_0\,G[\rho_0,\Chem_0]\,\rho_0=\mathcal{O}(\varepsilon)
\end{equation}
This condition is met when the system approaches a stationary solution. In this regime, the fine-scale effects due to the stochastic nature of the cell re-polarisation process may become determinative for the shape and stability of the cell profile. However, such a regime can not describe long-range migration. Instead, directed collective movement is properly captured by the evolution of the system toward a travelling-wave solution~\eqref{eq:travelling wave ansatz}. A diffusion scaling (see~\Cref{App:DiffLimit}) may be applied to study the relaxation dynamics of solutions to~\eqref{macroscopic mass balance hydrodynamics} towards the travelling wave~\eqref{eq:travelling wave ansatz} at long times. This would capture how the stochastic nature of the re-polarisation process affects the travelling speed and cell density profile during the migration dynamics. However, we find that this correction remains negligible except for relatively high values of $\varepsilon$ (see~\Cref{fig:study error for growth driven invasion} and discussion in~\Cref{sec: result proliferation-driven invasion}).

\subsection{Leading-order dynamics: drift-force balance}\label{sec:multiple scale}
When proliferation acts on a slower timescale --\emph{i.e.}, $\phi_0=0$, the standard asymptotic expansion~\eqref{eq:general_asymptotic_expansion_eps} breaks. As detailed in Appendix~\ref{sec:fail_hydro_limit}, in fact, when growth enters as an $\mathcal{O}(\varepsilon)$ perturbation, the leading-order macroscopic dynamics reduces to a purely advective conservation law
\begin{equation}\label{eqref: hydro leading order no prol}
    \frac{\partial \rho_0}{\partial \tau}+\nabla_{\bxi}\cdot\left[\bU[\rho_0,\Chem_0]\rho_0\right]=0, 
\end{equation}
subject to the no-flux boundary conditions~\eqref{eq: no flux_macro leading order} and with advection velocity $\bU$ as defined in~\eqref{eq: definition total cell velocity}. As for the previous case, the effective velocity incorporates the contribution of the interaction forces and the leading-order drift $\bV_T$~\eqref{UT}, which is coupled to the chemoattractant dynamics~\eqref{macroscopic chemo}. Since~\eqref{eqref: hydro leading order no prol} is a pure drift equation subject to no-flux conditions, at the leading order, the total cell mass is conserved. However, this is because proliferation only enters when looking at higher-order corrections to the leading-order dynamics, forcing the correction term to carry a non-zero mass. While negligible for short transients, this cumulative growth dominates on long time scales, invalidating the standard expansion. To avoid this issue, we employ a multiple-scale analysis, introducing a slow time variable to properly capture the long-term modulation of the migration front by cell proliferation.

\subsubsection{Method of multiple scales}
To account for the long-term effects of proliferation on the invasion dynamics described -- at leading order -- by \eqref{eqref: hydro leading order no prol}, we employ the method of multiple scales~\cite{bender_multiple-scale_1999,celora2026chemotaxiscellaggregatesmorphology}. Specifically, we introduce the slow time variable $\sigma=\varepsilon\tau$. Treating $\tau$ and $\sigma$ as independent variables, we seek solutions to \eqref{macroscopic mass balance hydrodynamics} of the form:
\begin{equation}\label{multiple scale redefinition}
\rho=\rho(\tau,\sigma,\bxi),\quad \mathcal{A}=\mathcal{A}(\tau,\sigma,\bxi).
\end{equation}
By the chain rule, the temporal derivative transforms as
\begin{equation}
    \frac{\partial}{\partial \tau} \hookrightarrow \frac{\partial}{\partial \tau}+\varepsilon \frac{\partial}{\partial \sigma}\,.\label{eq:chainrulemultiplescales}
\end{equation}
Substituting~\eqref{multiple scale redefinition}-\eqref{eq:chainrulemultiplescales} into~\eqref{macroscopic mass balance hydrodynamics} and~\eqref{macroscopic chemo}, and considering the standard asymptotic expansion of $\rho$ and $\Chem$ in powers of $\varepsilon$, as
\begin{equation}\label{eq:general_asymptotic_expansion_MS}
\varphi(\tau,\sigma,\bxi)=\sum_{n=0}^\infty \varepsilon^n\varphi_n(\tau,\sigma,\bxi)\,,\quad \varphi\in\left\{\rho,\Chem\right\}\,,
\end{equation}
we recover, at leading order, the same equation describing the purely advective dynamics for the cell density:
\begin{subequations}\label{eq:leading order multiple scales}
    \begin{align}
  \dfrac{\partial\rho_0}{\partial \tau}+\nabla_{\bxi} \cdot \left(\rho_0\bU[\rho_0,\Chem_0]\right)=0,\label{eq:multiple scales rho leading order}
\end{align} 
where $\bU$ is defined by~\eqref{eq: definition total cell velocity} and the chemoattractant distribution $\Chem_0$ evolves according to 
\begin{align}
   \dfrac{1}{\tau_{\Chem}}\frac{\partial\mathcal{A}_0}{\partial \tau} =\nabla^2_{\bxi}\mathcal{A}_0-\beta_\Chem[\rho_0,\Chem_0]\mathcal{A}_0.\label{eq:multiple scales A leading order}
\end{align}
\end{subequations}
Eqs.~\eqref{eq:leading order multiple scales} do not directly depend on the slow time scales $\sigma$, which can therefore be treated as a control parameter: $\rho_0=\rho_0(\tau,\bxi;\sigma)$ and $\mathcal{A}_0=\mathcal{A}_0(\tau,\bxi;\sigma)$. Assuming the no flux boundary conditions~\eqref{eq: no flux_macro leading order} on $\partial \Omega$, we can integrate~\eqref{eq:multiple scales rho leading order} on $\Omega$ finding that the mass 
\begin{equation}\label{def:N0}
N_0:=\int_\Omega\rho_0(\tau,\bxi;\sigma)d\bxi
\end{equation}
is conserved on the fast timescale, but its evolution is determined by the slow time scale $\sigma$, {\it i.e.,} ${N_0=N_0(\sigma)}$, which is yet to be determined. As mentioned above, we are interested in the regime where \eqref{eq:leading order multiple scales} converges to a travelling-wave solution of the form \eqref{eq:travelling wave ansatz}. Unlike in the drift-proliferation balance case, where the leading-order dynamics converged to the same solution independent of the initial conditions, the long-term behaviour of~\eqref{eqref: hydro leading order no prol} may depend on the initial condition via the conservation of the total mass. As a result, the migration speed $\vec{c}_0$ and the travelling-wave profile of the solution could depend on the slow timescale $\sigma$ via the total mass $N_0$:
\begin{equation}
    \lim_{\tau\to\infty} \varphi_0(\tau,\vec{\xi};\sigma)=\bar{\varphi}_0(\vec{\xi}-\vec{c_0}(N_0)\tau;N_0),\quad \varphi_0=\rho_0,\Chem_0\,.\label{eq:multiple scale TW}
\end{equation}
Hence, in this regime, proliferation does not directly drive migration, but rather, it mediates it by shaping the speed and spatial structure of the migrating front. 

\paragraph{Dynamics of the total cell number.} To obtain a closed-form solution to the problem~\eqref{eq:leading order multiple scales} and determine the evolution of $N_0$, we have to study the dynamics of the first-order correction $\rho_1$. Substituting~\eqref{multiple scale redefinition}-\eqref{eq:general_asymptotic_expansion_MS} into~\eqref{macroscopic mass balance hydrodynamics} and retaining the $\mathcal{O}(\varepsilon)$-terms, we obtain
\begin{equation}
    \dfrac{\partial\rho_1}{\partial \tau} + \nabla_{\bxi} \cdot \Big[\rho_1\bU[\rho_0,\Chem_0]+\rho_0\bV_T'[\Chem_0]\Chem_1+\rho_0\Lambda \bmW[\rho_1]\Big]=\nabla_{\bxi} \cdot \vec{Q}_0+\phi_1G[\rho_0,\Chem_0]\rho_0-\frac{\partial \rho_0}{\partial \sigma}\,,\label{first order correction multiple scales}
\end{equation}
where $\bU$ is defined in~\eqref{eq: definition total cell velocity}, $\bV'_T$ indicates the derivative of $\bV_T$ with respect to $\Chem$, and
\begin{equation}
\vec{Q}_0:=\nabla_{\bxi} \cdot(\bD_T[\Chem_0]\rho_0) - \rho_0\bU[\rho_0,\Chem_0]\nabla_{\bxi}\cdot \bV_T[\Chem_0]-\rho_0 \dfrac{\partial \bV_T[\Chem_0]}{\partial \tau}.
\end{equation}
The dynamics of $N_0$ on the slow timescale is then determined by integrating~\eqref{first order correction multiple scales} over $\Omega$ to obtain
\begin{equation}\label{eq:N1}
\frac{\partial N_1}{\partial \tau}= \phi_1\int_{\Omega} G[\rho_0,\mathcal{A}_0]\rho_0\,d\bxi-\frac{dN_0}{d\sigma}\,,
\end{equation}
where the flux terms in~\eqref{first order correction multiple scales} vanish once we integrate in space due to the no-flux boundary conditions. We can integrate~\eqref{eq:N1} explicitly with respect to the fast time $\tau$ to find
\begin{equation}
    N_1(\tau;\sigma)=N_1(0;\sigma)-\int_0^\tau \left[\dfrac{dN_0}{d\sigma}-\phi_1\int_\Omega G[\rho_0(\tau',\bxi;\sigma),\mathcal{A}_0(\tau',\bxi;\sigma)]\rho_0(\tau',\bxi;\sigma)\,d\bxi\,\right]d\tau'.\label{eq:N1explicit}
\end{equation}
As detailed in Appendix~\ref{app:travelling wave convergence}, formalising the convergence of the solution to the leading-order problem~\eqref{eq:leading order multiple scales} to a travelling wave of the form~\eqref{eq:multiple scale TW}, and under additional assumptions on the properties of the proliferation rate function $G$~\eqref{eq:proliferation kernel}, we can show that there exists a time $\tau_1\in(0,\infty)$ and two constants $b,\lambda>0$, independent of $\tau$, such that
\begin{equation}
N_1(\tau;\sigma) = n_1(\tau;\sigma)-\left[\dfrac{dN_0}{d\sigma}-\bar{r}(\sigma)\right](\tau-\tau_1), \quad \forall\tau\geq\tau_1,\label{eq:estimate for secular term}
\end{equation}
where
\[
\bar{r}(\sigma):=\phi_1\int_\Omega G[\bar{\rho}_0,\bar{\mathcal{A}}_0]\,\bar{\rho}_0\,d\bxi\,
\]
and the function $n_1(\tau; \sigma)$ satisfies 
\begin{equation}\label{eq: convergence condition main}
    |n_1|\leq\Gamma+ \phi_1\int_0^{\tau_1} \int_\Omega \left|G[\rho_0(\tau',\bxi;\sigma),\Chem_0(\tau',\bxi;\sigma)]\rho_0(\tau',\bxi;\sigma)\right|\,d\bxi\, d\tau' +b(1-e^{-\lambda (\tau-\tau_1)}),\quad \forall \tau\geq\tau_1\,,
\end{equation}
where $\Gamma:=\left|N_1(0,\sigma)\right|+\left|\tau_1\frac{dN_0}{d\sigma}\right|$. We note that while the constants $b$ and $\lambda$ are independent of time $\tau$, they may depend on the slow timescale $\sigma$. Given~\eqref{eq: convergence condition main}, we can conclude that the function $n_1$ is bounded in $\tau$. Therefore, to prevent %the emergence of a secular term that would cause 
$N_1$~\eqref{eq:estimate for secular term} to grow unboundedly as $\tau \to \infty$, we must ensure that the secular term, which is linear in $\tau$, vanishes. This yields the following solvability condition 
\begin{equation}\label{eq:evolution mass slow time}
    \frac{d N_0}{d\sigma}=\bar{r}(\sigma),
\end{equation}
which defines the slow-time evolution of the leading-order total cell mass $N_0$~\eqref{def:N0}. 
\begin{oss*}
The constants $\tau_1$, $b$ and $\lambda$ that determine the asymptotic behaviour of $n_1$~\eqref{eq: convergence condition main} do not influence the leading-order evolution of the total cell mass~\eqref{eq:evolution mass slow time}. Yet, they determine the accuracy of~\eqref{eq:evolution mass slow time} in describing the evolution of the total cell mass. In general, the faster the system converges to the corresponding travelling-wave profile -- \emph{i.e.,} the larger $\lambda$-- the more accurate~\eqref{eq:evolution mass slow time} becomes. In addition to the influence of the arbitrarily chosen initial conditions, the approximation may also lose accuracy in regimes where the travelling-wave profiles sharply change with $N_0$. In these regimes, transient dynamics can drive more significant perturbations to the mass that can not be captured by the leading-order approximation. These are typically associated with discontinuous morphological transitions in the cell density profile such as the one observed in~\cite{ford_pattern_2024,celora2026chemotaxiscellaggregatesmorphology}. However, this is not the case for the examples analysed in this work (see~\Cref{sec: result drift-driven invasion}).
\end{oss*}

To summarise, while~\eqref{eqref: hydro leading order no prol} captures the initial transient dynamics, it fails to capture the long-time behaviour of the system. This is because, over longer periods, the increase in cell mass due to proliferation has a non-negligible impact on the invasion process. The method of multiple scales allow to properly account for the cumulative effect of growth over longer time, offering a macroscopic description of invasion in slowly proliferating cell populations.

\section{Long-range migration of cell streams and clumps}\label{sec:example}
\begin{figure}[htb]
    \centering
\begin{subfigure}{0.00\textwidth}
     \centering
        \captionlistentry{}
        \label{fig:simulations TW with growth A}
\end{subfigure}
\begin{subfigure}{0.975\textwidth}
    \includegraphics[width=\textwidth]{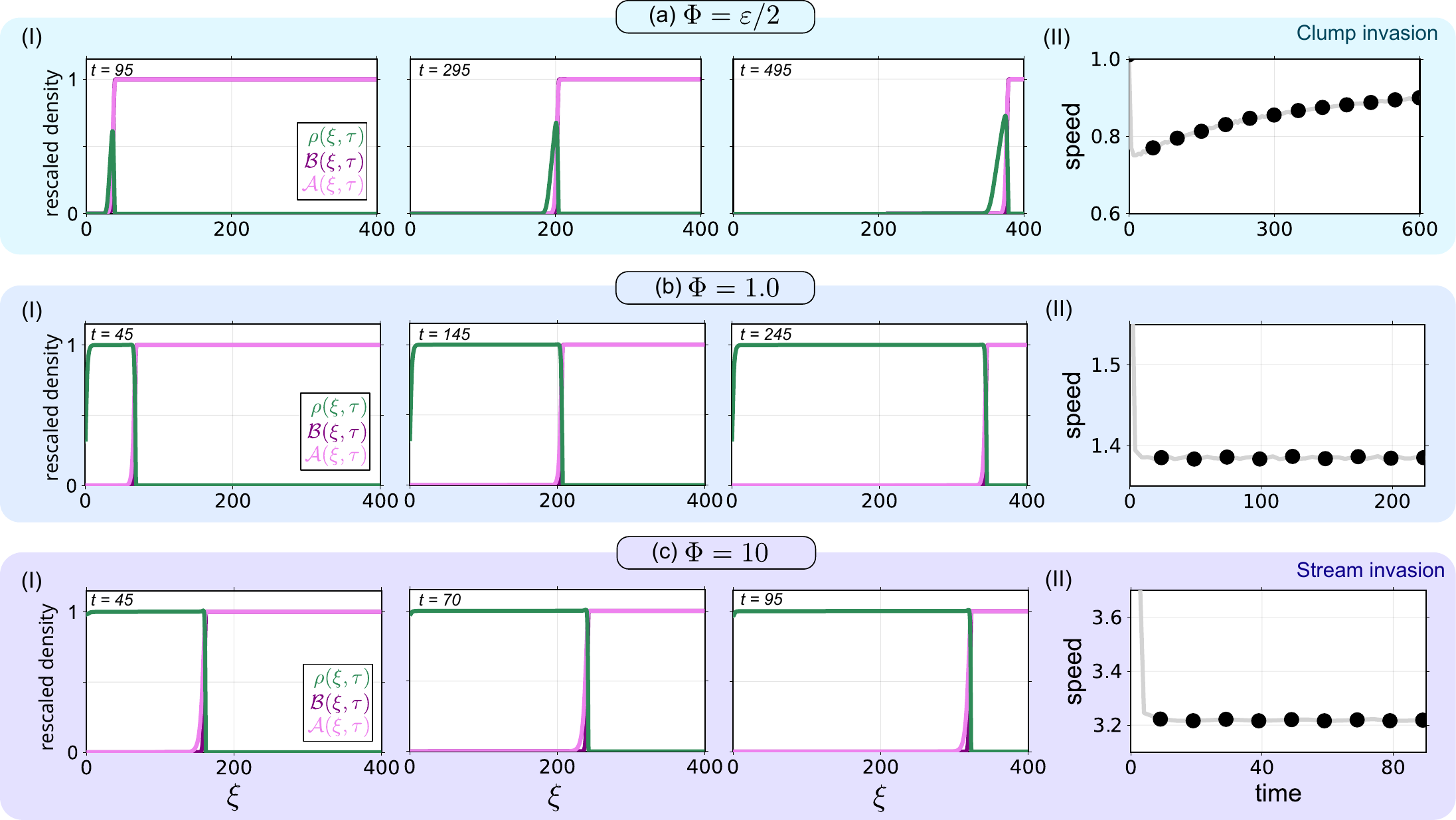}
     \captionlistentry{}
                \label{fig:simulations TW with growth B}
\end{subfigure}
\begin{subfigure}{0.00\textwidth}
     \centering
        \captionlistentry{}
                \label{fig:simulations TW with growth C}
\end{subfigure}
\vspace{-5mm}
    \caption{{\bf Impact of cell proliferation on front propagation dynamics.} Numerical simulation of the first-order hydrodynamic theory~\eqref{macroscopic mass balance hydrodynamics}-\eqref{eq: general no flux_macro} for $\varepsilon=0.01$ and different values of the proliferation rate $\Phi$: (a) $\Phi=0.005$, (b) $\Phi=1.0$ and (c) $\Phi=10$. Other model parameters are set to the values in~\Cref{tab:model parameters}. ({\sf I}) Spatial profile of the cell density $\rho$ (green curve), chemoattractant source $\mathcal{B}$ (purple curve), and chemoattractant $\Chem$ (pink curve) at different times during the simulations. Time increases from left to right. ({\sf II}) Estimates of the front propagation speed, where the front is defined as the spatial location at which $\Chem(\tau,\xi_f(\tau))=0.5$. }
    \label{fig:simulations TW with growth}
\end{figure}

We apply the macroscopic models derived in~\Cref{sec:macroscopic limit} to systematically study the impact of proliferation on the collective chemotaxis of interacting cells.~In particular, we test the implications of the two asymptotic timescale regimes identified on the dynamics of collective cell invasion. To this end, we propose a minimal setup by which the interplay between self-generated chemotaxis and short-range volume exclusion is sufficient to drive long-range migration of cell clumps and streams depending on the rate of cell proliferation (\Cref{fig:simulations TW with growth}). 

While the framework developed in Section~\ref{sec:macroscopic limit} applies to arbitrary spatial dimensions, for simplicity, we apply it to a simplified pseudo-two-dimensional setting. Specifically, we consider a population of cells migrating over a flat substrate (Figure~\ref{fig:schematicA}) and assume the solution is homogeneous along the y-direction, so that the problem reduces to one spatial dimension on the half-line: $\vec{\xi}=\xi \vec{e}_x$ with $\xi>0$. Under these assumptions, we can compute the moments of the turning operator~\eqref{eq:turning operator} explicitly (see Eqs.~\eqref{eq: advection speed turning kernel}-\eqref{eq: diffusion tensor aerotaxis} in~\Cref{app:numerical simulations}). As standard in previous studies of chemotaxis~\cite{tweedy_self-generated_2016,bhattacharjee_chemotactic_2021,ucar_self-generated_2025}, we assume a logarithmic shape for the chemoattractant sensing function 
\begin{equation}
    S(\Chem)=\log\left(\frac{\mathcal{A}}{\mathcal{A}+a_+}\right),\label{eq:definition sensing function}
\end{equation}
where the constant $a_+\geq0$ is the dissociation constant for binding of the chemoattractant to the receptors on the cell surface; receptors get saturated above this concentration, preventing cells from measuring gradients~\cite{tweedy_self-generated_2016}. 

Similarly to~\cite{ford_pattern_2024,celora2026chemotaxiscellaggregatesmorphology}, we assume that cells shape chemoattractant gradients via feeding over a second chemical species, here denoted by $\mathcal{B}$, which acts as a source for the chemoattractant $\Chem$, setting 
\begin{equation}
\beta_\Chem[\rho,\Chem]=-\mathcal{B}[\rho]+1\label{eq:clump def consumption}
\end{equation}
in~\eqref{eq:multiple scales A leading order} and allowing $\mathcal{B}$ to follow its own reaction dynamics
\begin{equation}
    \frac{1}{\tau_\mathcal{B}}\frac{\partial \mathcal{B}}{\partial \tau}=\partial_{\xi\xi}\mathcal{B}+\Phi_\mathcal{B}G_{\mathcal{B}}(\mathcal{B})\mathcal{B}-\beta_{\mathcal{B}}\rho\mathcal{B}\,.\label{eq:clum food}
\end{equation}
Here the constants $\Phi_{\mathcal{B}}>0$ and $\beta_{\mathcal{B}}>0$ represent the non-dimensional rates of chemoattractant source production and food consumption by cells, respectively. The function $G_{\mathcal{B}}$ captures the reaction dynamics of $\mathcal{B}$ in the absence of cells; we choose it to model a \emph{strong Allee effect}, namely $G_{\mathcal{B}}(\mathcal{B})=(1-\mathcal{B})(1-\mathcal{B}/\mathcal{B}^*)$, where the constant $\mathcal{B}^*\in(0,1)$ indicates the threshold limit below which $\mathcal{B}$ naturally decays. In contrast, for initial concentrations above the threshold $\mathcal{B}^*$, the population $\mathcal{B}$ grows up to its carrying capacity, which is here normalised to $1$. The interplay between the spatially distributed chemoattractant source $\mathcal{B}$ ensures exponential decay of the chemoattractant profile $\mathcal{A}$ at the rear of the invading front. Combined with the logarithmic sensing function~\eqref{eq:definition sensing function}, the exponential decay of $\mathcal{A}$ ensures a constant chemotaxis bias at the rear of the invading front/clump~\cite{celora2026nonlineartheorychemotacticfronts}. This is sufficient to sustain travelling-wave solutions~\eqref{eq:travelling wave ansatz} even in the absence of proliferation.

Finally, while our framework is valid for a general interaction kernel of the form~\eqref{eq: definition W}, for the numerical results, we focus on the regime in which cells interact only via volume exclusion, \emph{i.e.}, setting $\kappa=0$ and $\nu=1$ in~\eqref{eq: definition W}. Under these assumptions, the velocity contribution from cell-cell interactions simplifies to $\bmW[\rho] = -\nu\partial_\xi\rho$,
effectively introducing a density-dependent diffusion term into the conservation equation~\eqref{macroscopic mass balance hydrodynamics}, which encapsulates the macroscopic effects of short-range volume exclusion. We further consider volume-filling effects for cell proliferation by assuming a logistic form for the proliferation rate 
\begin{equation}
    G(\rho,\Chem)=(1-\rho).\label{eq:G definition general}
\end{equation}

\subsection{Summary of the model}
Under the above assumptions, the first-order hydrodynamic theory derived in~\Cref{sec:hyperbolic scaling standard} takes the form
{\allowdisplaybreaks
\begin{subequations}\label{app:1D_model}
    \begin{align}
         \dfrac{1}{\tau_{\Chem}}\dfrac{\partial \Chem}{\partial \tau}&=\partial_{\xi\xi}\,\Chem+(\mathcal{B}-1)\Chem\,,&\quad \xi>0,\ t>0,\label{app:chemoattractant dynamics}\\[0.2cm]
          \frac{1}{\tau_\mathcal{B}}\frac{\partial \mathcal{B}}{\partial \tau}&=\partial_{\xi\xi}\mathcal{B}+\Phi_\mathcal{B}(1-\mathcal{B})\left(1-\frac{\mathcal{B}}{\mathcal{B}^*}\right)\mathcal{B}-\beta_{\mathcal{B}}\rho\mathcal{B},&\quad \xi>0,\ t>0,\label{app:source dynamics}\\[0.2cm]
            \frac{\partial \rho}{\partial \tau}&= -\partial_{\xi}\left[ \rho U[\rho,\Chem]+\varepsilon \rho \tilde{U}[\rho,\Chem] \right]+(\phi_0+\varepsilon\phi_1)(1-\rho)\rho\,,&\quad \xi>0,\ t>0,\label{app:dynamics cell density}
    \end{align}
    where the leading- and first-order corrections to the cell velocity in~\eqref{app:dynamics cell density} are defined as:
    \begin{align}
        U&:= V_T - \Lambda\nu \partial_\xi\rho,\\
        \tilde{U}&:= -\partial_\xi D_T + \frac{\partial V_T}{\partial \tau}+U\partial_\xi V_T- D_T\partial_\xi \ln\rho\,.
    \end{align}
\end{subequations}
}
The scalar functions $V_T$ and $D_T$ are associated with the moments of the turning kernel~\eqref{UT}-\eqref{VT}. The mean drift $\bV_T$ reads
\begin{equation}
    \bV_T= V_T\hat{\vec{e}}_x,\quad V_T(\partial_\xi S(\Chem))=\frac{I_1(\alpha\partial_\xi S(\Chem))}{I_0(\alpha\partial_\xi S(\Chem))} \,, \label{eq: advection speed turning kernel}
\end{equation}
while the diffusion tensor $\bD_T$ is
\begin{equation}\label{eq: diffusion tensor aerotaxis}
\bD_T=D_T(\hat{\vec{e}}_x\otimes\hat{\vec{e}}_x), \quad   D_T(\partial_\xi S(\Chem))=\frac{1}{2}\left(1+\frac{I_2(\alpha\partial_\xi S(\Chem))}{I_0(\alpha\partial_\xi S(\Chem))}\right)-V_T^2, %u_\parallel^2, 
     \end{equation}
where $I_i$ are the modified \emph{Bessel} functions of the first kind. The relation between the sensing function $S$ and the local chemoattractant gradients is defined as in~\eqref{eq:definition sensing function}. The list of non-dimensional parameters in~\eqref{app:1D_model} is given in~\Cref{tab:model parameters}.

\begin{table}[htb]
    \centering
    \begin{tabular}{c||p{125mm}|c}
    \toprule
         & Description  & Value(s) \\
         \midrule
         $\alpha$ & parameter mediating the strength of the orientation bias in the turning operator~\eqref{eq:turning operator} &10\\
         $a_+$& normalised dissociation constant for binding of chemoattractant with the receptors on the cell surface~\eqref{eq:definition sensing function}&1\\
         $\tau_\Chem$& characteristic relaxation time-scale for the chemoattractant dynamics~\eqref{app:chemoattractant dynamics}&1\\
          $\Phi_{\mathcal{B}}$& rescaled proliferation rate of the chemoattractant source~\eqref{app:chemoattractant dynamics}&1\\
         $\tau_\mathcal{B}$& characteristic relaxation time-scale for the chemoattractant source dynamics~\eqref{app:source dynamics}&1\\
         $\beta_{\mathcal{B}}$ &consumption rate of the chemoattractant source per cell~\eqref{app:source dynamics}&4 \\
         $\mathcal{B}^*$ &Allee constant for the chemoattractant source~\eqref{app:source dynamics}&0.1 \\
         $\nu$&normalised cell motility parameter associated with volume exclusion& 1\\
         $\Lambda$&dimensionless parameter associated with viscous spreading (\Cref{sec:macroscopic limit})&1\\
         $\sigma$ & initial spread of the cell population~\eqref{app:initial conditions} & 10\\
         $N_0$ &initial cell mass~\eqref{app:initial conditions}&$\approx 3$\\
         \bottomrule
    \end{tabular}
    \caption{List of parameters in the macroscopic invasion model~\eqref{app:1D_model}-\eqref{app:initial conditions} along with their default value used to generate Figures~\ref{fig:simulations TW with growth}-\ref{fig:clump invasion dynamics with prol}, unless differently specified.}
    \label{tab:model parameters}
\end{table}

We close~\eqref{app:1D_model} by imposing appropriate boundary and initial conditions. At the boundary $\xi=0$, we assume a no-flux boundary condition for all species:
\begin{align}\label{app: BC inlet}
    \partial_\xi \mathcal{A}=\partial_\xi \mathcal{B}= \rho(U+\varepsilon\tilde{U})=0, \quad \xi=0,\  t>0,
\end{align}
while in the far-field we assume that all species converge to a constant:
\begin{equation}\label{app: BC far-field}
    \lim_{\xi\to\infty} \mathcal{A}=\lim_{\xi\to\infty}\mathcal{B}=1, \quad \lim_{\xi\to\infty}\rho=0.
\end{equation}
Note that~\eqref{app: BC far-field} implies that cells have an infinite reservoir of chemoattractant and that the total cell density remains finite. We choose initial conditions that are consistent with~\eqref{app: BC inlet}-\eqref{app: BC far-field}; specifically:
\begin{equation}
    \rho(\xi,0)= N_0\sqrt{\frac{2}{\pi\sigma^2}}\exp\left(-\frac{\xi^2}{2\sigma^2}\right),\quad \Chem(\xi,0)\equiv 1,\quad \mathcal{B}(\xi,0)\equiv 1,\label{app:initial conditions}
\end{equation}
where $\sigma$ captures the initial spread of the population, while $N_0$ corresponds to the total initial rescaled density of cells in the domain. 

\subsection{Numerical Results}

We solve~\eqref{app:1D_model}-\eqref{app:initial conditions} numerically. Details on the numerical scheme used and boundary and initial conditions are given in~\Cref{app:numerical simulations}. The predicted invasion dynamics for increasing values of the proliferation rate $\Phi$ are shown in Figure~\ref{fig:simulations TW with growth}. As we sweep $\Phi$ across different asymptotic regimes, we find that the dynamics of invasion substantially change. For $\Phi\sim\mathcal{O}(\varepsilon)$, we find that cells migrate as a clump that slowly increases in size. Changes in the size of the clump are coupled to its migration speed, which slowly increases over time. In contrast, once $\Phi\gtrsim \mathcal{O}(1)$, the cells invade the space as a continuous stream. In this regime, the migration speed of the invading front quickly converges to a constant value, and the cell distribution relaxes to a travelling-wave profile that spreads from the inlet, where the concentration of cells is below saturation as a result of the no-flux boundary conditions. 

The numerical results presented in~\Cref{fig:simulations TW with growth} support the use of different asymptotic regimes to describe clump vs stream invasion, as identified in Section~\ref{sec:macroscopic limit}. We now aim to apply the theory developed in Sections~\ref{sec:macroscopic limit} %and~\ref{sec:fail_hydro_limit} 
to understand the pattern of invasion observed in Figures~\ref{fig:simulations TW with growth A} and~\ref{fig:simulations TW with growth C}. We first consider the regime in which proliferation is one of the mechanisms driving long-range migrations using the macroscopic theory developed in~\Cref{sec:stand_drift_prol}. We then move on to consider the less studied case in which proliferation is much slower via the theory outlined in~\Cref{sec:multiple scale}. 

\subsubsection{Stream migration: proliferation-driven invasion}
\label{sec: result proliferation-driven invasion}

We apply the leading-order hydrodynamic model~\eqref{eqref: hydro leading order with prol}-\eqref{eqref:hydro leading order with prol A} to understand proliferation-driven invasion (as in Figures~\ref{fig:simulations TW with growth B}-\ref{fig:simulations TW with growth C}). This regime describes continuous, stream-like invasion, in which self-generated chemotaxis at the leading edge is balanced by local proliferation in the trailing population, thereby sustaining persistent front propagation. A classic biological example is neural crest cell migration during embryogenesis, where migrating cell streams establish self-generated chemoattractant gradients while proliferation replenishes the migrating population, enabling long-range invasion~\cite{mclennan2012multiscale,landman2007mathematical}.  Mechanistically, this framework is structurally analogous to the porous-Fisher-KPP model, where degenerate diffusion accounts for volume exclusion, combined with a chemotaxis flux whose speed depends non-linearly on chemical gradients~\eqref{eq: advection speed turning kernel}.

\begin{figure}[htb]
\begin{subfigure}{0.0\textwidth}
    \captionlistentry{}
    \label{fig:study error for growth driven invasion A}
\end{subfigure}
\begin{subfigure}{0.0\textwidth}
    \captionlistentry{}
    \label{fig:study error for growth driven invasion B}
\end{subfigure}
    \centering
    \includegraphics[width=\linewidth]{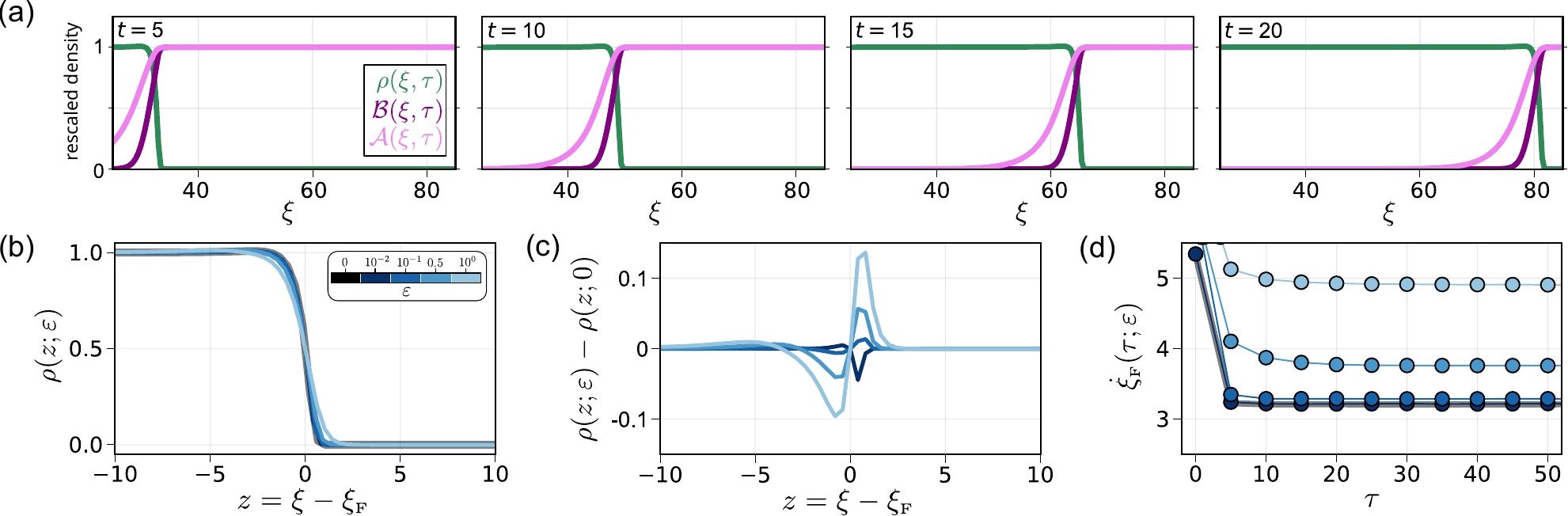}
    \begin{subfigure}{0.0\textwidth}
    \captionlistentry{}
    \label{fig:study error for growth driven invasion C}
\end{subfigure}
\begin{subfigure}{0.0\textwidth}
    \captionlistentry{}
    \label{fig:study error for growth driven invasion D}
\end{subfigure}
\vspace{-6mm}
    \caption{{\bf Stream migration.} Study of the numerical solution to~\eqref{macroscopic mass balance hydrodynamics}-\eqref{eq: general no flux_macro} with~\eqref{eq:clump def consumption}-\eqref{eq:clum food} for increasing value of $\varepsilon$ in the case of proliferation-driven invasion ($\phi_0=10$). Other model parameters are set to the values in~\Cref{tab:model parameters}. (a) Snapshot of the solution profiles of the cell density $\rho$ (green curve), chemoattractant source $\mathcal{B}$ (purple curve), and chemoattractant $\Chem$ (pink curve) predicted by the leading-order model~\eqref{eqref: hydro leading order with prol}-\eqref{eqref:hydro leading order with prol A} at distinct time points.  Time increases from left to right. (b) Comparison of the asymptotic travelling-wave profile $\rho_\varepsilon$. The origin of the reference frame corresponds to the location of the front $\xi_F$, \emph{i.e.}, where $\rho(0;\varepsilon)=0.5$. (c) Local error in the approximation of the solution by the leading-order solution $\rho(z;0)$. (d) Difference in the evolution of the front speed. }
    \label{fig:study error for growth driven invasion}
\end{figure}

The results of the simulation of~\eqref{eqref: hydro leading order with prol}-\eqref{eqref:hydro leading order with prol A} for $\phi_0=10$ (see Eq.~\eqref{eq:definition Phi}) are shown in~\Cref{fig:study error for growth driven invasion A}. We find that the solution quickly relaxes to a travelling-wave profile. In contrast to the prediction of the standard porous-Fisher-KPP model~\cite{simpson_fisherkpp-type_2024}, the cell density profile is non-monotonic. At the rear of the wave, the profile is determined by proliferation and the density saturates to its carrying capacity (here normalised to unity). Near the front, a non-monotonic peak in the cell density emerges. This feature is common in travelling-wave solutions arising from the coupling of growth and self-generated chemotaxis ~\cite{narla_traveling-wave_2021} and originates from the decrease of the chemotaxis bias near the front compared to the bulk region.

In general, the leading-order model agrees with the hydrodynamic theory~\eqref{macroscopic mass balance hydrodynamics}-\eqref{eq: general no flux_macro}. In particular, it recapitulates the asymptotic travelling wave regime both qualitatively and quantitatively in the small-$\varepsilon$ regime (Figures~\ref{fig:study error for growth driven invasion B}-\ref{fig:study error for growth driven invasion D}) up to values of $\varepsilon \sim \mathcal{O}(1)$. In particular, the error in the travel profile remains negligible across the range of $\varepsilon$ considered. The largest discrepancies are confined to the invasion front, where diffusion smooths the interface and progressively reduces the height of the density peak (Figures~\ref{fig:study error for growth driven invasion B}-\ref{fig:study error for growth driven invasion C}). Consequently, as $\varepsilon$ increases, the travelling-wave profile approaches a sigmoidal shape similar to that predicted by the Fisher–KPP model. Despite the negligible changes in the travelling profile, we find a significant increase in the travelling-wave speed (\Cref{fig:study error for growth driven invasion D}) when $\varepsilon\sim\mathcal{O}(1)$. Nonetheless, we conclude that the leading-order model~\eqref{eqref: hydro leading order with prol}-\eqref{eqref:hydro leading order with prol A} remains an accurate approximation to the first-order hydrodynamic theory~\eqref{macroscopic mass balance hydrodynamics}-\eqref{eq: general no flux_macro} well beyond its regime of validity.

To summarise, our numerical simulations reveal that rapidly proliferating cell populations migrate as streams whose migration speed is directly influenced by the rate at which they proliferate.

\subsubsection{Clump migration: mass-mediated invasion}\label{sec: result drift-driven invasion}
We now consider the role of proliferation in mediating the migration of cell clumps or aggregates rather than continuous streams. This pattern occurs in {\it E. coli} chemotactic bands \cite{saragosti2010mathematical}, where interplay between chemotaxis and growth constraints dictates cluster geometry \cite{mittal2003motility}. Beyond bacteria, cell invasion often proceeds via detached, self-sustaining aggregates~\cite{ford_pattern_2024,panigrahi_intermittent_2025}--classic in vivo example being the zebrafish posterior lateral line, guided by a self-generated chemokine gradient \cite{li2004chemokine}. Similar collective dynamics also occur in {\it Dictyostelium discoideum} via oscillatory cAMP signaling \cite{gregor2010onset}.

\begin{figure}[htb]
    \centering
    \begin{subfigure}{0\textwidth}
    \captionlistentry{}
    \label{fig:clump invasion dynamics no prol A}
    \end{subfigure}
        \begin{subfigure}{0\textwidth}
    \captionlistentry{}
    \label{fig:clump invasion dynamics no prol B}
    \end{subfigure}
        \begin{subfigure}{0\textwidth}
    \captionlistentry{}
    \label{fig:clump invasion dynamics no prol C}
    \end{subfigure}
        \begin{subfigure}{0\textwidth}
    \captionlistentry{}
    \label{fig:clump invasion dynamics no prol D}
    \end{subfigure}
    \includegraphics[width=\linewidth]{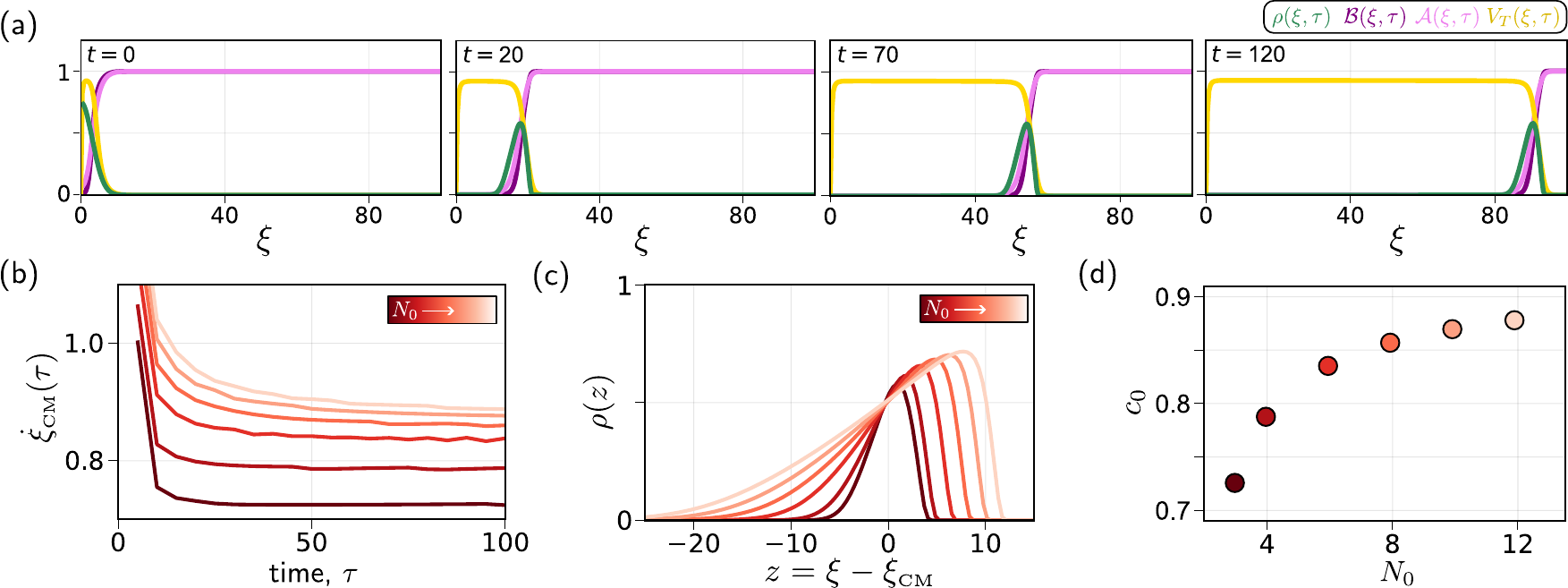}
\caption{{\bf Clump migration on the fast time scale.} Numerical simulation of the leading-order hydrodynamic limit~\eqref{eq:leading order multiple scales},~\eqref{eq:clump def consumption}-\eqref{eq:clum food}, for different choices of the initial mass $N_0$. Other model parameters are set to the values in~\Cref{tab:model parameters}. (a) Snapshot of the solution profiles of the cell density $\rho$ (green curve), chemoattractant source $\mathcal{B}$ (purple curve), chemoattractant $\Chem$ (pink curve), and first moment of the turning kernel $V_T$~\eqref{eq: advection speed turning kernel} (yellow curve) at distinct time points for $N_0=3$. Time increases from left to right. (b) Comparison of the time evolution of the speed at which the centre of mass of the clump ($\xi_{\text{\tiny CM}}:=\int_0^L\xi\rho(\tau,\xi) \d\xi/N_0$) moves for different values of $N_0\in\left\{3,4,6,8,10,12\right\}$. In all cases, the speed converges to a constant value $c_0$ corresponding to the speed of the travelling aggregate. (c) Comparison of the asymptotic cell density profile $\rho$ for different values of $N_0$. The spatial variable is shifted so that the origin corresponds to the centre of mass of the moving clump. (d) Relation between the travelling speed $c_0$ and the initial clump mass. }    
\label{fig:clump invasion dynamics no prol}
\end{figure}

A characteristic simulation of~\eqref{eq:leading order multiple scales} is illustrated in Figure~\ref{fig:clump invasion dynamics no prol A}. As the cell population $\rho$ migrates into the domain, it feeds on $\mathcal{B}$, locally decreasing its concentration below $\mathcal{B}^*$. This creates a permanent spatial asymmetry between the region where the cell aggregate has previously been, and the region ahead where the concentration of food $\mathcal{B}$ remains high. This asymmetry allows for the formation of a persistent, self-generated chemoattractant gradient that travels with the cell aggregate~(\Cref{fig:clump invasion dynamics no prol A}). After a short transient, the solution converges to an asymmetric travelling-wave profile with constant speed (Figure \ref{fig:clump invasion dynamics no prol B}). We note that a travelling-wave profile develops for all choices of the initial aggregate mass $N_0$ (Figure \ref{fig:clump invasion dynamics no prol C}), provided it exceeds a critical threshold (result not shown). This threshold represents the minimum amount of droplet mass required to locally push the concentration of the source $\mathcal{B}$ below the Allee threshold $\mathcal{B}^*$. Once travelling-wave solutions are established, both the propagation speed and the travelling wave morphology become uniquely determined by the clump mass $N_0$~(\Cref{fig:clump invasion dynamics no prol B}-\ref{fig:clump invasion dynamics no prol D}), which is set by the initial condition. As $N_0$ increases, the clump experiences a stronger chemotactic pulling force as it generates steeper gradients. This eventually saturates, as cells reach their maximal speed -- here normalised to unity. Yet, the speed of the clump saturates below this critical value. This is due to the spatial heterogeneity of cell behaviour across the moving aggregate, whereby cells at the leading front move more slowly. The difference in the speed of cells in the rear and front of the clump increases with $N_0$, resulting in a more asymmetric profile. The steep change in the chemotactic speed with increasing $N_0$ tightly compresses the front of the aggregate. Conversely, at the rear of the wave, where the food is exhausted, the cells move at a constant speed (see yellow curve in~\Cref{fig:clump invasion dynamics no prol A}) and the morphology of the clump is shaped by volume-exclusion.
\begin{figure}[htb]
    \centering
    \begin{subfigure}{0\textwidth}
    \captionlistentry{}
    \label{fig:clump invasion dynamics with prol A}
    \end{subfigure}
        \begin{subfigure}{0\textwidth}
    \captionlistentry{}
    \label{fig:clump invasion dynamics with prol B}
    \end{subfigure}
        \begin{subfigure}{0\textwidth}
    \captionlistentry{}
    \label{fig:clump invasion dynamics with prol C}
    \end{subfigure}
        \begin{subfigure}{0\textwidth}
    \captionlistentry{}
    \label{fig:clump invasion dynamics with prol D}
    \end{subfigure}
            \begin{subfigure}{0\textwidth}
    \captionlistentry{}
    \label{fig:clump invasion dynamics with prol E}
    \end{subfigure}
    \includegraphics[width=\linewidth]{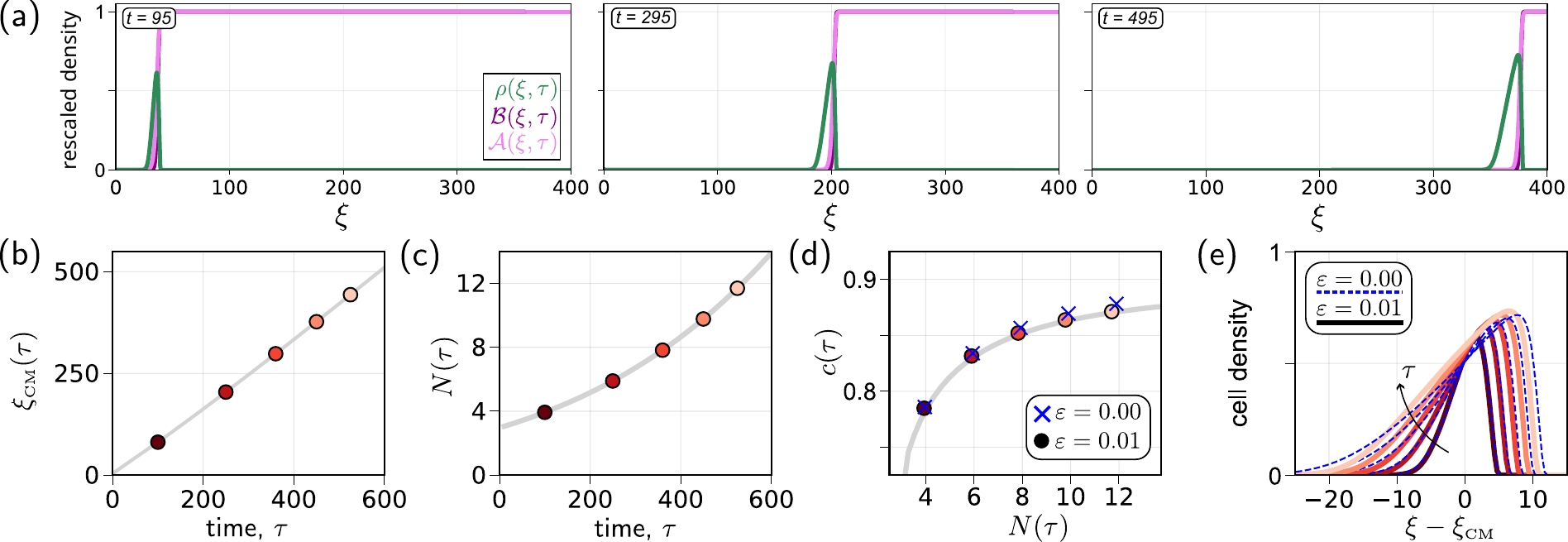}
\caption{{\bf Clump migration on the slow time scale.} Comparison of the hydrodynamic limit~\eqref{macroscopic mass balance hydrodynamics} and multiple-scale analysis for $\varepsilon=0.005$. Other model parameters are set to the values in~\Cref{tab:model parameters}. (a) Snapshot of the spatial profile of the cell density $\rho$ (green curve), chemoattractant source $\mathcal{B}$ (purple curve), and chemoattractant $\Chem$ (pink curve) at different times during the simulations. Same as in~\Cref{fig:simulations TW with growth A}. Time increases from left to right. (b) Time evolution of the location of the clump centre of mass $\xi_{\text{\tiny CM}}$ (as defined in~\Cref{fig:clump invasion dynamics no prol}). (c) Time evolution of the clump mass. (d) Estimated relationship between the clump mass and speed. (e) Comparison of the cell density profile $\rho$ at different time points in the reference frame of the centre of mass. The metrics for the first-order approximation are extracted from the same numerical simulations used to produce~\eqref{fig:simulations TW with growth A}.}  %defined as the speed at which its centre of mass $\bar{\xi}_\rho$ moves, \emph{i.e.}, $c:=\d\xi_\text{\tiny CM}/\d\tau$.  
\label{fig:clump invasion dynamics with prol}
\end{figure}

The results of the numerical simulations show that the long-term cell distribution and migration speed predicted by~\eqref{eq:leading order multiple scales}, coupled with~\eqref{eq:clump def consumption}-\eqref{eq:clum food}, depend dynamically on the evolution of the total cell mass (\Cref{fig:clump invasion dynamics no prol}). This implies that any perturbation driven by slow proliferation to the clump mass would not relax; rather, its effect accumulates over time. As a result, over extended timescales, the mass-conserving leading-order approximation~\eqref{eqref: hydro leading order no prol} fails to predict the macroscopic clump dynamics (Figures~\ref{fig:clump invasion dynamics with prol A}). We can capture this dynamics via the multiple scale framework developed in~\Cref{sec:multiple scale}. This successfully resolves the invasion dynamics of cell clumps (Figures~\ref{fig:clump invasion dynamics with prol D}-\ref{fig:clump invasion dynamics with prol E}), describing it as a slow transition through a continuous family of travelling-wave solutions parametrised by the instantaneous mass $N(\tau)$. As time progresses, the clump advances~(\Cref{fig:clump invasion dynamics with prol B}) while its mass $N(\tau)$ grows continuously~(\Cref{fig:clump invasion dynamics with prol C}). Simultaneously, the clump dynamically adjusts its migration speed $c(\tau)$~(\Cref{fig:clump invasion dynamics with prol D}) and spatial structure~(\Cref{fig:clump invasion dynamics with prol E}) to match the travelling-wave profile dictated by its current mass.

\section*{Discussion and conclusion}\label{sec:discussion} 

In this work, we introduced a novel multiscale framework of collective cell invasion able to explain different modes of invasion observed across diverse biological contexts: cell streams and clumps. Specifically, we describe cells as proliferative, active, and interacting particles, whose drift speed is biased by self-generated gradients in an external chemoattractant field. We systematically derived a hydrodynamic theory of cell invasion by taking a macroscopic limit of the underlying mesoscopic description of the particle ensemble. To properly account for the different relative timescales in the problem, we employ techniques drawn from both the classical kinetic theory of active particles and multiple-scale analysis. Our approach differs from existing upscaling techniques that assume that migration and proliferation adopt a diffusive scaling. In contrast, our framework successfully captures transport-dominated invasion dynamics in which invasion and proliferation operate on decoupled timescales. As a result, the appropriate macroscopic description is determined not only by the underlying kinetic model, but also by the relative timescale associated with the biological mechanisms included. This is crucial to account for the distinctive dynamics of stream and clump invasion.

Our findings reveal two distinct modes of macroscopic collective invasion depending on how cell kinetics couple with the migration forces. When proliferation and chemotactic movement evolve on comparable timescales, a classical leading-order hydrodynamic theory can be employed to derive a continuity equation for the total cell density. Under these conditions, we observed the emergence of travelling streams~(Figures~\ref{fig:simulations TW with growth B}-\ref{fig:simulations TW with growth C}). Biologically, this is common in neural crest development~\cite{mclennan2012multiscale,landman2007mathematical,simpson2007cell}, where proliferation or influx of cells from boundaries is sufficient to fill the space behind the invading front. 

In contrast, when proliferation is much slower than chemotactic movement, the leading-order hydrodynamic limit only captures the transient dynamics. We recover the invasion dynamics via a multiple-scale analysis, which properly accounts for the impact of slow changes in the cell mass on the cell invasion dynamics. In this regime, we observe the long-range migration of cohesive cell clumps. While proliferation is not necessary for invasion~(\Cref{fig:clump invasion dynamics no prol}), it shapes the dynamics by influencing both the clump morphology and invasion speed~(\Cref{fig:clump invasion dynamics no prol}-\ref{fig:clump invasion dynamics with prol}). Our multi-scale framework frames the long-term migration as a slow, quasi-static transition through a continuous family of travelling-wave solutions parametrised by the instantaneous mass $N(\tau)$. Hence, even when decoupled from fast migration dynamics, cell proliferation plays a critical role in dictating the long-term sustainability, structure, and velocity of the invading population. Consequently, the higher-order corrections act as a selection mechanism that continuously adjusts both the clump morphology and its collective invasion velocity as proliferation slowly increases the cell mass. This is consistent with previous analyses of phenomenological models of clump invasion~\cite{celora2026chemotaxiscellaggregatesmorphology} and reflects dynamics observed in complex biological systems, such as in amoeboid motion~\cite{ford_pattern_2024} and development of the zebrafish posterior lateral line~\cite{li2004chemokine,wu_collective_2025}, where a cluster of cells migrates cohesively guided by self-generated chemoattractant gradients.

Our multiscale framework opens up several directions for future investigation. For simplicity, here we assumed that the cell turning rate is independent of environmental stimuli. More generally, however, $\mu$ may depend on $\mathcal{A}$ or its spatial gradients. For instance, in the presence of strong chemoattractant gradients, cells frequently exhibit increased persistence and reorient less frequently, effectively reducing the turning rate to stabilize their directional flight. Such environment-dependent turning mechanisms have been widely considered in kinetic models of directed migration (see, e.g.,~\cite{block1983adaptation,othmer2002diffusion}) and could be integrated into our hydrodynamic scaling to explore how adaptive persistence modulates clump stability and front propagation. Furthermore, while our model assumes a constant speed for the cell, it can be naturally extended to incorporate heterogeneous speed distributions or speed-jump processes within the turning dynamics~\cite{erban2004individual,loy2020kinetic,conte2026novel}. Non-local effects could also be included to capture long-range sensing of environmental cues over a finite spatial neighbourhood~\cite{loy2020kinetic,conte2023non,conte2022multi}. Moreover, it would be interesting to extend the asymptotic analysis beyond the leading-order approximation by deriving the dynamics of the higher-order corrections, with the aim of further exploring the connection between the present hydrodynamic limit and the classical parabolic diffusion regime.

For simplicity, we here considered the invasion dynamics of a homogeneous cell population. Yet, many biological systems achieve robust collective migration through specialised subpopulations, such as leader–follower~\cite{mclennan2012multiscale,Sara_Bernardi_2021_Kinetic,jewell2026cellcelladhesionsustainextended} or consumer–sensor~\cite{ucar_self-generated_2025,li2004chemokine,celora2026nonlineartheorychemotacticfronts,watts2026coupled} organisations. Incorporating such heterogeneity would substantially broaden the biological applicability of the proposed framework while also addressing interesting mathematical challenges. In particular, distinct cell subpopulations may naturally evolve on different migration or proliferation timescales. The coupling of distinct kinetic equations increases the range of possible asymptotic scalings, especially where heterogeneity yields asymmetric scaling limits. 

Overall, our work contributes to the theoretical understanding of collective cell invasion by showing how the same kinetic model can give rise to different macroscopic theories. These distinct descriptions arise from the separation of different biological timescales, which is reflected theoretically in the need for different asymptotic descriptions. Ultimately, these theoretical advances provide a rigorous framework to decipher and predict the multiscale nature of cell streams and clumps, offering valuable insights into complex biological processes ranging from developmental morphogenesis to collective cancer invasion.

\paragraph{Acknowledgments}
The authors thank Philip Maini and Mohit Dalwadi for helpful feedback and discussions. GLC acknowledges financial support from a Hooke Research Fellowship. MC acknowledges support from the National Group of Mathematical Physics (GNFM-INdAM) (CUP: E5324001950001) and the European Union - Next Generation EU, Mission 4, Component 1 (CUP: E13C24002380006). MC also acknowledges partial support by the State Research Agency of the Spanish Ministry of Science and FEDER-EU, project PID2022-137228OB-I00 and by Modeling Nature Research Unit, Grant QUAL21-011 funded by Consejería de Universidad, Investigaci\'on e Innovaci\'on (Junta de Andalucía). For the purpose of Open Access, the authors have applied a CC BY public copyright licence to any Author Accepted Manuscript (AAM) version arising from this submission. 

\appendix
\section{Additional result: hydrodynamic theory}
In this section, we provide additional calculations that complement the derivation of the hydrodynamic theory of collective cell migration presented in \Cref{sec:macroscopic limit}.

\subsection{Derivation of the correction term to cell flux~\eqref{eq:flux fperp}}\label{app:fperp_derivation}
We provide the detailed calculations to determine the first-order correction $f^{\perp}\in\langle T(\hv)\rangle^{\perp}$ appearing in the Chapman--Enskog expansion \eqref{eq:chapman expansion} of the distribution function $f$. This is then used to approximate the flux term in~\eqref{eq:density_advection_chapman_expansion}. We start by substituting the expansion~\eqref{eq:chapman expansion} into the rescaled kinetic equation~\eqref{transport_eq_rescale_hyp} to obtain
\begin{equation}\label{app: expression kinetic equation with chapman expansion}
\begin{split}
    \varepsilon\frac{\partial(\rho T[\Chem]+ \varepsilon f^{\perp})}{\partial \tau}+\varepsilon\nabla_{\bxi}\cdot[(\hv+\Lambda \bmW[\rho])(\rho T[\Chem]+\varepsilon f^{\perp})]=\varepsilon\mathcal{L}f^{\perp}+\varepsilon\Phi \mathcal{P}(\rho T[\Chem]+\varepsilon f^{\perp})\,.
    \end{split}
\end{equation}
By collecting terms at leading order, the equation yields
\begin{equation}
\begin{split}
   \mathcal{L}\fperp&=\frac{\partial (\rho T[\Chem])}{\partial \tau}+\nabla_{\bxi}\cdot\Big[(\hv+\Lambda \bmW[\rho])\rho T[\Chem]\Big]-\mathcal{P}\,\rho T[\Chem]\\[0.2cm]
   &=\rho\frac{\partial T[\Chem]}{\partial \tau} +T[\Chem]\Big[\left(\hv-\bV_T[\Chem]\right)\cdot\nabla_{\bxi} \rho-\rho\nabla_{\bxi}\cdot \bV_T[\Chem]\Big] +\rho(\hv+\Lambda \bmW[\rho])\cdot \nabla_{\bxi} T[\Chem]
   \end{split}
\end{equation}
where we have utilized the properties of the turning operator. To apply the pseudo-inverse of $\mathcal{L}$ and guarantee the existence of the solution $f^{\perp}$, we first verify the solvability condition dictated by the Fredholm alternative, namely
\[
\int_{\mathbb{S}^{n-1}}\mathcal{L}\fperp\,d\hv=0
\]
which holds true thanks to the properties of the turning kernel $T[\Chem]$. Hence, we can apply the pseudo-inverse and obtain
\begin{equation}\label{app:approximate expression for f perp}
\begin{split}
   \fperp&=-\rho\frac{\partial T[\Chem]}{\partial \tau} -T[\Chem]\Big[\left(\hv-\bV_T[\Chem]\right)\cdot\nabla_{\bxi} \rho-\rho\nabla_{\bxi}\cdot \bV_T[\Chem]\Big] -\rho(\hv+\Lambda \bmW[\rho])\cdot \nabla_{\bxi} T\,.
   \end{split}
\end{equation}
Following \cite{hillen2006m5} and keeping implicit the dependence of $T$ and $\bV_T$ on the chemoattractant $\Chem$ for convenience of notation, we multiply~\eqref{app:approximate expression for f perp} by $\hv$ and integrate over the orientation space $\mathbb{S}^{n-1}$. This yields the first-order flux contribution: 

\begin{align}
&\int\limits_{\mathbb{S}^{n-1}} \hv \fperp d\hv\\
&=-\rho \frac{\partial}{\partial \tau}\int\limits_{\mathbb{S}^{n-1}} \hv T d\hv-\int\limits_{\mathbb{S}^{n-1}}\hv T\left[(\hv-\bV_T)\cdot\nabla_{\bxi} \rho-\rho\nabla_{\bxi}\cdot \bV_T\right]d\hv -\int\limits_{\mathbb{S}^{n-1}}\hv\rho(\hv+\Lambda \bmW[\rho])\cdot \nabla_{\bxi} T d\hv\notag\\[0.1cm]
&=-\rho \frac{\partial \bV_T}{\partial \tau}-\!\!\!\!\int\limits_{\mathbb{S}^{n-1}}\hv\otimes\hv\, T \,d\hv\,\nabla_{\bxi} \rho+\bV_T\otimes \bV_T\nabla_{\bxi} \rho+\rho\bV_T\nabla_{\bxi}\cdot \bV_T-\!\!\!\!\int\limits_{\mathbb{S}^{n-1}}\hv\otimes \hv \, \nabla_{\bxi} T \,d\hv\,\rho-\rho\Lambda \bmW[\rho]\nabla_{\bxi}\cdot \bV_T\notag\\[0.15cm]
&\hspace{1.5cm}\Longrightarrow \int\limits_{\mathbb{S}^{n-1}} \hv \fperp d\hv=-\left(\rho \frac{\partial \bV_T}{\partial \tau}+\nabla_{\bxi}\cdot(\rho\bD_T)+\rho\left(\bV_T+\Lambda \bmW[\rho]\right)\nabla_{\bxi}\cdot \bV_T \right)\,.\label{app: intermediate result for fperp}
\end{align}
Finally, substituting the first-order correction flux~\eqref{app: intermediate result for fperp} into the macroscopic conservation law~\eqref{eq:density_advection_chapman_expansion}, we arrive at the closed drift-diffusion equation for the macro-density $\rho$, accurate up to order $\mathcal{O}(\varepsilon)$:
\begin{equation}
    \frac{\partial \rho}{\partial \tau}=\nabla_{\bxi}\cdot\left[ \varepsilon\nabla_{\bxi}\cdot(\bD_T\rho) -\left(\left(\bV_T+\Lambda \bmW[\rho]\right)+\varepsilon \tilde{\bV}\right)\rho\right]+\Phi\, G[\rho, \Chem] \rho
\end{equation}
where $\bD_T=\bD_T[\Chem]$, $\bV_T=\bV_T[\Chem]$, and  
\begin{equation}
    \tilde{\bV}=\tilde{\bV}[\Chem]:=-\left(\bV_T[\Chem]+\Lambda \bmW[\rho]\right)\nabla_{\bxi}\cdot \bV_T[\Chem]-\frac{\partial \bV_T[\Chem]}{\partial \tau}\,.
\end{equation}
\subsection{Beyond initial transient dynamics: proliferation negligible}\label{sec:fail_hydro_limit}

Considering the scaled transport equation~\eqref{eq:f rescaled} under the hyperbolic scaling~\eqref{eq:hyperbolic scaling} with dominant proliferation ($\Phi \sim \mathcal{O}(\varepsilon)$), the kinetic equation takes the form
\begin{equation}\label{transport_eq_rescale_multiScale}
\varepsilon\dfrac{\partial}{\partial \tau} f(\tau,\bxi,\hv) +\varepsilon  \nabla_{\bxi} \cdot (\left(\hv+\Lambda\bmW\right) f(\tau,\bxi,\hv))=\mathcal{L}f(\tau,\bxi,\hv) + \varepsilon^2\phi_1 \mathcal{P}f(\tau,\bxi,\hv).
\end{equation}
Applying the Chapman–Enskog procedure~\eqref{eq:chapman expansion} described in Section \ref{sec:hyperbolic scaling standard} to~\eqref{transport_eq_rescale_multiScale} yields a closed macroscopic equation for the density $\rho$. Formally expanding the cell density in powers of $\varepsilon$~\eqref{eq:general_asymptotic_expansion_eps}, %i.e., $\rho \sim \rho_0 + \varepsilon\rho_1$, 
we find that the leading-order $\rho_0$ and first-order correction $\rho_1$ satisfy
\begin{subequations}\label{hyp_lim_correction}
\begin{align}
\dfrac{\partial\rho_0}{\partial \tau}+\nabla_{\bxi} \cdot \left[\rho_0\bU[\rho_0,\Chem_0]\right]&=0 , \label{eq leading order wrong multiplescales}\\[0.2cm]
\dfrac{\partial\rho_1}{\partial \tau}
\!+\! \nabla_{\bxi} \cdot \Big[\rho_1\bU[\rho_0,\Chem_0]+\rho_0\left(\bV_T'[\Chem_0]\Chem_1+\Lambda \bmW[\rho_1]\right)\Big]
&= \nabla_{\bxi}\cdot\left[\nabla_{\bxi} \cdot(\bD_T[\Chem_0]\rho_0) +\rho_0\tilde{\bV}[\Chem_0]\right]\label{eq first order wrong multiplescales}\\[0.15cm]
&\quad+ \phi_1 G[\rho_0,\Chem_0]\rho_0,
\end{align}
\end{subequations}
with $\bU$ defined in~\eqref{eq: definition total cell velocity}, $\tilde{\bV}$ in~\eqref{eq: definition V tilde}, and $\bV'_T$ indicating the derivative of $\bV_T$ with respect to $\Chem$. As expected in a classical hyperbolic regime where advection dominates, the leading-order dynamics~\eqref{eq leading order wrong multiplescales} is purely transport-driven, whereas diffusive and proliferative mechanisms only emerge at order $\mathcal{O}(\varepsilon)$. Proliferation therefore results in the correction $f^{\perp}$ in~\eqref{eq:chapman expansion} carrying a non-zero mass. Although this expansion is formally consistent, the presence of proliferation affects the uniform validity of the asymptotic approximation in time. Integrating~\eqref{eq first order wrong multiplescales} over space shows that the proliferative term generates a non-vanishing source inducing a secular growth of size $\varepsilon \tau$ in the total mass.  While negligible for $\tau = \mathcal{O}(1)$, this term becomes comparable to the leading-order terms on time intervals of order $\tau = \mathcal{O}(\varepsilon^{-1})$.
 
Thus, system~\eqref{hyp_lim_correction} offers an accurate macroscopic description only in the absence of proliferation, where cell motion is strictly governed by the mean velocity field $\bU$, with higher-order terms merely accounting for local velocity reorientations. The system~\eqref{hyp_lim_correction} would therefore provide the correct macroscopic dynamics only in the absence of proliferation. In that case, cell motion is governed solely by directed transport along the macroscopic mean velocity field $\bU$, and higher-order corrections introduce dispersive and diffusive effects associated with local reorientations of the microscopic velocity distribution. To overcome this limitation and properly capture the interplay between slow cell proliferation and fast migration dynamics, in Section~\ref{sec:multiple scale} we adopt the method of multiple scales~\cite{bender_multiple-scale_1999}.

\subsection{Derivation of the estimates~\eqref{eq:estimate for secular term}}
\label{app:travelling wave convergence}
In this Appendix, we detail the derivation of~\eqref{eq:estimate for secular term} and the assumptions under which this holds. For simplicity, we introduce the vectorial notation for the solution $$\vec{y}_i(\tau,\bxi;\sigma)=[\rho_i(\tau,\bxi;\sigma),\mathcal{A}_i(\tau,\bxi;\sigma)] \in L^2_\infty(\mathbb{R}_+\times\Omega\times\mathbb{R}_+;\mathbb{R}_+^2).$$ We assume that for $\tau\gg1$ the function $\vec{y}_0$ exponentially converges to the travelling-wave solution with finite mass
\begin{equation}
\bar{\vec{y}}_0=\bar{\vec{y}}_0(\bxi-\vec{c}\tau,\sigma)\in L_\infty(\Omega\times\mathbb{R}_+)\bigcap L_1(\Omega\times\mathbb{R}_+)\,.\tag{C1}
\end{equation}
Specifically, we assume that
\begin{equation}
    \exists\tau_1>0, \text{ s.t. } \forall\tau>\tau_1,\quad \|\vec{y}_0-\bar{\vec{y}}_0\|_\infty+\|\rho_0-\bar{\rho}_0\|_1\leq Y e^{-\lambda (\tau-\tau_1)}, \label{eq:condition convergence}\tag{C2}
\end{equation}
for a given $\lambda>0$, which is the rate of convergence of the solution to $\bar{\vec{y}}_0$, and constant $Y>0$. Note that all constants are in principle dependent on the slow timescale $\sigma$, but to simplify the notation we do not write this explicitly. 

We further assume that the function $G$ in~\eqref{eq:f rescaled} is smooth, so that there exists $M>0$ such that
\begin{equation}
    \|G[\vec{y}^1]-G[\vec{y}^2]\|_\infty\leq M\|\vec{y}^2-\vec{y}^2\|_{\infty}, \quad \forall \vec{y}^1,\vec{y}^2\in R\subset\mathbb{R}^2,\label{eq:condition Lipschitz on G}\tag{C3}
\end{equation}
where $R$ is a bounded domain, and, given the convergence properties~\eqref{eq:condition convergence}, it is generally sufficient that $[-\|\bar{\rho}_0\|_\infty-Y,\|\bar{\rho}_0\|_\infty+Y]\times[-\|\bar{\mathcal{A}}_0\|_\infty-A,\|\bar{\mathcal{A}}_0\|_\infty+A]\subset R$, for a given constant $A>0$. 

Considering now Eq.~\eqref{eq:N1explicit} and separating the contribution for $\tau<\tau_1$ and $\tau\geq\tau_1$, we can write
\begin{equation}
    N_1(\tau,\sigma)=N_1(\tau_1,\sigma)-\left[\frac{dN_0}{dt}-\bar{r}(\sigma)\right](\tau-\tau_1)+\int_{\tau_1}^{\tau}\int_\Omega \left[G[\rho_0,\Chem_0]\rho_0-G[\bar{\rho}_0,\bar{\Chem}_0]\bar{\rho}_0\right]d\vec{\xi}d\tau'\,.
\end{equation}
We proceed by showing that the contribution to the last term is bounded as $\tau\to\infty$. In particular, we have that
\begin{equation}\label{eq: bound derivation 1}
\begin{aligned}
    \left|\int_{\tau_1}^{\tau}\int_\Omega \Big[G[\vec{y}_0]\rho_0-G[\bar{\vec{y}}_0]\bar{\rho}_0\Big]d\vec{\xi}d\tau'\right|&\leq \int_{\tau_1}^{\tau}\int_\Omega \left|G[\vec{y}_0]-G[\bar{\vec{y}}_0]\right|\rho_0+|G[\bar{\vec{y}}_0]|\,\|\rho_0-\bar{\rho}_0\|_1d\vec{\xi}d\tau'\\
    &\leq \int_{\tau_1}^\tau \|G[\vec{y}_0]-G[\bar{\vec{y}}_0]\|_\infty N_0 + G_{\max}\|\rho_0-\bar{\rho}_0\|_1d\tau',
    \end{aligned}
\end{equation}
where we define $G_{\max}=\max\limits_{\bxi\in\Omega}|G[\bar{\vec{y}}_0(\bxi)]|$.
Using the additional assumptions~\eqref{eq:condition Lipschitz on G}, we can further bound the last term in~\eqref{eq: bound derivation 1} to obtain
\begin{equation}
\begin{aligned}
    \int_{\tau_1}^\tau \|G[\vec{y}_0]-G[\bar{\vec{y}}_0]\|_\infty N_0 + G_{\max}\|\rho_0-\bar{\rho}_0\|_1d\tau'&\leq\int_{\tau_1}^\tau MN_0\|\vec{y}_0-\bar{\vec{y}}_0\|_{\infty}+G_{\max} \|\rho_0-\bar{\rho}_0\|_1d\tau'\\
    &\leq\max\left\{MN_0,G_{\max}\right\}Y\int_{\tau_1}^{\tau} e^{-\lambda(\tau'-\tau_1)}d\tau'\\
    &=b\left(1-e^{-\lambda(\tau-\tau_1)}\right)
    \end{aligned}
\end{equation}
where $$b=\dfrac{\max\left\{MN_0,G_{\max}\right\}Y}{\lambda}$$
which is generally a function of the slow variable $\sigma$. This leads to the expression~\eqref{eq: convergence condition main} in the main text.

\section{Beyond hydrodynamic theories of collective cell migrations}
In this work, we have focused on the derivation of macroscopic models of long-range collective migration based on the hyperbolic scaling~\eqref{eq:hyperbolic scaling}, which is defined in the asymptotic regime $\text{Kn}=\varepsilon$, with ${\text{Kn}:=v/(l\mu)}$. However, different physical balances are possible for~\eqref{eq:f rescaled} (see \Cref{tab:different limits}), depending on the size of the average cell-flux $\rho\bv_c$~\eqref{eq:macroscopic cell flux}. In particular, when $\rho\bv_c$ is an $\mathcal{O}(\varepsilon)$-quantity, the dynamics of cell movement occur on a much slower timescale, which is comparable to that at which diffusive corrections to the orientation-jump processes occur. As a result, upscaling of~\eqref{eq:f rescaled} via diffusion limits is required to capture the dynamics of collective migration. 

\begin{table}[h!] 
    \centering
    \begin{tabular}{c|l|c}
    \toprule[1.5pt]
         &\parbox{55mm}{\centering$\tilde{t}=1/\varepsilon^2\mu$} & $\tilde{t}=1/\varepsilon\mu$ \\[3pt]
          \hline\addlinespace[3pt]
        $g\sim\varepsilon^2\mu$&   \parbox{50mm}{\centering diffusion limit (\Cref{App:DiffLimit})\\ $ t_{v}\ll\tilde{t}\sim t_P\sim t_{RM}$} &  \parbox{80mm}{\centering multiple scales hydrodynamic limit (\Cref{sec:multiple scale})\\ $\tilde{t}\sim t_{v}\ll t_P \sim t_{RM}$} \\[7pt]       
         \hline\addlinespace[3pt]
          $g\sim\varepsilon\mu$& \parbox{50mm}{\centering proliferation dominates}  & \parbox{80mm}{\centering hydrodynamic limit (\Cref{sec:stand_drift_prol})\\ $\tilde{t}\sim t_{v}\sim t_P\ll t_{RM}$} \\[8pt]
        \bottomrule[1.5pt]
    \end{tabular}
    \caption{Summary of the different physical balances admissible for Eq.~\eqref{eq:f rescaled}. We indicate with the subscripts (P) proliferation, ($v$) directed motion, and (RM) random motion, and $\tilde{t}$ indicates the macroscopic selected timescale.}
    \label{tab:different limits}
\end{table}
\subsection{Diffusive (or parabolic) limit}
\label{App:DiffLimit}
The diffusive limit of a transport equation of the form of Eq.~\eqref{eq:f rescaled} has been extensively studied (\emph{e.g.}, see \cite{othmer2000diffusion,othmer2002diffusion}). For completeness, we here summarise only the main results and refer the reader to the literature for a detailed analysis.

We consider the rescaled transport equation~\eqref{eq:f rescaled} and  we assume now that the macroscopic time scale is comparable to the characteristic time of random motion, {\it i.e.,}
\begin{equation}\label{eq: diffusion scaling}
\tilde{t}\sim t_{RM}=\mu\frac{l^2}{v^2}\gg t_{v},
\end{equation}
as we are in the so-called high frequency regime. The scaling~\eqref{eq: diffusion scaling} leads to a diffusion limit as $\text{St}=\varepsilon$. To allow for diffusion to dominate over directed cell motion, we further assume that:
\begin{equation}
    \Lambda=\varepsilon \tilde{\Lambda}, \quad \text{and } \bV_T\sim\mathcal{O}(\varepsilon).\label{eq:scaling for diffusion balance}
\end{equation}
Since the scaling for $\bV_T$ depends on the chemoattractant, the second condition assumes that the chemoattractant distribution has specific properties. For our choice of turning kernel~\eqref{eq:turning operator}, this implies the factor $\alpha|\nabla S(\Chem)|\sim\mathcal{O}(\varepsilon)$. This is the case if the gradients in the sensing function $S$ are shallow or if the variance of the angular distribution $\alpha^{-1}$ is large, which would correspond to the case of weak chemotactic bias. Here, we consider this second scenario, setting $\alpha=\tilde{\alpha}\varepsilon$ into~\eqref{eq:turning operator} and expanding in powers of $\varepsilon$ yields
\begin{equation}\label{eq:expansion turning kernel small variability}
T[\Chem](\hv)=\frac{1}{|\mathbb{S}^{n-1}|}+\frac{\tilde{\alpha}\varepsilon}{|\mathbb{S}^{n-1}|}\frac{\langle \hv,\nabla S(\Chem)\rangle}{|\nabla S(\Chem)|}+\frac{\tilde{\alpha}^2\varepsilon^2}{2|\mathbb{S}^{n-1}|}\left(\frac{\langle \hv,\nabla S(\Chem)\rangle^2}{|\nabla S(\Chem)|^2}-\frac{1}{n}\right)+\mathcal{O}(\varepsilon^3),
\end{equation}
which is analogous to the perturbative structure of the turning kernel adopted by~\cite{hillen2006m5}. Here, bias due to the chemoattractant gradient only enters as a correction to the diffusive random motion of the cell, provided that the gradient $|\nabla S|\lesssim\mathcal{O}(1)$.

In the diffusion regime (see Table~\ref{tab:different limits}) we have $\Phi\sim\mathcal{O}(1)$ and, under~\eqref{eq:scaling for diffusion balance}, the kinetic transport equation~\eqref{eq:f rescaled} takes the form
\begin{equation}\label{transport_eq_rescale_par}
\varepsilon^2\dfrac{\partial}{\partial \tau} f(\tau,\bxi,\hv) +\varepsilon \nabla_{\bxi}\cdot \left( \left(\hv+\varepsilon\tilde\Lambda \bmW[\rho]\right)f(\tau,\bxi,\hv)\right)=\mathcal{L}f(\tau,\bxi,\hv) +\varepsilon^2\phi_0\mathcal{P}f(\tau,\bxi,\hv).
\end{equation}%
We employ a classical asymptotic expansion in power of $\varepsilon$ for $f$ and $\Chem$ that captures higher-order corrections
\begin{equation}\label{eq:general_asymptotic_expansion}
    \begin{aligned}
    f(\tau,\bxi,\hv)&= f_0(\tau,\bxi,\hv)+\varepsilon f_1(\tau,\bxi,\hv) +\varepsilon^2 f_2(\tau,\bxi,\hv)+\mathcal{O}(\varepsilon^3),\\
    \Chem(\tau,\bxi)&=\Chem_0(\tau,\bxi)+\varepsilon\Chem_1(\tau,\bxi)+\varepsilon^2\Chem_2(\tau,\bxi)+\mathcal{O}(\varepsilon^3)\,.
    \end{aligned}
\end{equation}
Since the turning kernel $T$ depends on the macroscopic variables $\Chem$ and $\rho$, the expansions~\eqref{eq:general_asymptotic_expansion} implies the analogous expansion
\begin{equation}\label{T_exp_gen}
   \begin{aligned}
  & T[\Chem](\hv)=T_0[\Chem_0](\hv)+\varepsilon T_1[\Chem_0,\Chem_1](\hv)+\varepsilon^2 T_2[\Chem_0,\Chem_1,\Chem_2](\hv)+\mathcal{O}(\varepsilon^3)\\[0.2cm]
  \end{aligned}
\end{equation}
and similarly for its moments. Substituting~\eqref{eq:expansion turning kernel small variability} into~\eqref{condition_T_norm}, the physical constrain implies that the normalisation condition yields 
\begin{equation}\label{Ti_ass}
    \int\limits_{\mathbb{S}^{n-1}}T_i(\hv)d\hv=\delta_{i0}\,,\,\,\forall i=0,1,\ldots\,.
\end{equation}
The additional assumption~\eqref{eq:scaling for diffusion balance} on the first moment of the turning operator further requires that the leading-order macroscopic velocity vanishes, namely
\begin{equation}\label{zero_mean}
    \bV_T^0:=\int\limits_{\mathbb{S}^{n-1}}\hv T_0[\Chem_0](\hv) d\hv=0\,\qquad\text{for} \,\,\text{ a.e. }\, \bxi\in\Omega\,,\ \tau\ge0.
\end{equation}
This condition ensures that the directed component of the operator appears only as a first-order correction. 

To derive a macroscopic parabolic description of the evolution of the cell density $\rho$ from the underlying kinetic model, we substitute the asymptotic expansions~\eqref{eq:general_asymptotic_expansion}–\eqref{T_exp_gen} into the rescaled governing equation~\eqref{transport_eq_rescale_par} for the distribution function. We then collect terms of equal order in $\varepsilon$ to obtain the corresponding hierarchy of equations. At leading order, we find that $f_0$ is the equilibrium associated with the leading-order turning operator, namely
\begin{equation}\label{f0_eq}
    \mathcal{L}_0f_0(\tau,\bxi,\hv):=\left(\rho_0T_0[\Chem_0](\hv)-f_0(\hv)\right)=0 \Longrightarrow f_0(\tau,\bxi,\hv)=\rho_0(\tau,\bxi)T_0[\Chem_0](\hv).
\end{equation}
At order $\varepsilon$, we obtain
    \begin{equation}\label{eps1_par}
    \begin{split}
    \nabla_{\bxi} \cdot \left(\hv f_0(\tau,\bxi,\hv)\right)=\left(T_0[\Chem_0](\hv)\,\rho_1(\tau,\bxi)-f_1(\tau,\bxi,\hv)\right)+ T_1[\Chem_0,\Chem_1](\hv)\,\rho_0(\tau,\bxi)\,.
        \end{split}
\end{equation}
To determine $f_1$, we invert the leading-order turning operator $\mathcal{L}_0$. Using conditions~\eqref{Ti_ass} together with the assumption on the leading-order macroscopic velocity, we obtain
\begin{equation}\label{f1_eq}
    f_1(\tau,\bxi,\bv)=-\nabla_{\bxi}\cdot\left(\hv f_0(\tau,\bxi,\hv)\right)+T_1[\Chem_0,\Chem_1](\hv)\,\rho_0(\tau,\bxi)
\end{equation}
which in turn implies 
\begin{equation}\label{rho1_zero}
    \rho_1(\tau,\bxi)=0\,.
\end{equation}
Finally, substituting the expressions for $f_0$ and $f_1$ from~\eqref{f0_eq} and~\eqref{f1_eq} into the $\mathcal{O}(\varepsilon^2)$ equation in~\eqref{transport_eq_rescale_par}, and imposing the solvability condition required to invert $\mathcal{L}_0 f_2$, we obtain the macroscopic diffusion equation for $\rho_0$:
\begin{equation}\label{macro_diff_gen}
\dfrac{\partial \rho_0}{\partial \tau} -\nabla_{\bxi}\cdot\left[\nabla_{\bxi}\cdot (\,\bD_T^0[\Chem_0]\rho_0)\right]+\nabla_{\bxi}\cdot\left[\left(\bV_T^1[\Chem_0,\Chem_1]+\tilde\Lambda \bmW[\rho_0]\right)\rho_0\right]=\phi_0 G[\rho_0,\Chem_0]\rho_0
\end{equation}
Here, the advection velocity $\bV_T^1$ corresponds to the first-order correction of the first moment of the turning kernel, while the diffusion tensor $\bD_T^0$ is obtained from~\eqref{VT} applied to the leading-order term of the turning kernel expansion.

\begin{oss*}
When proliferation is neglected, the total mass of the system is conserved, namely:
\begin{equation}
\begin{split}
    N_0&=\int\limits_{\mathbb{S}^{n-1}}\int\limits_\Omega f(\tau,\bxi,\hv)d\bxi d\hv=\int\limits_{\mathbb{S}^{n-1}}\int\limits_\Omega (f_0(\tau,\bxi,\hv)+\varepsilon f_1(\tau,\bxi,\hv)+\mathcal{O}(\varepsilon^2))d\bxi d\hv\\[0.2cm]
    &=\int\limits_\Omega (\rho_0(\tau,\bxi)+\varepsilon \rho_1(\tau,\bxi)+\mathcal{O}(\varepsilon^2))\,d\bxi
\end{split}
\end{equation}
In this setting, it is natural to assume that the total mass is entirely captured by the leading-order term \cite{loy2020kinetic}, {\it i.e.,}
\begin{equation}\label{ass_rhoi}
 \rho_0(t,\x)=\rho(t,\x)\,\quad \text{and} \quad\,\rho_i(t,\x)=0\,,\quad a.e.\quad \x\in\Omega, \quad t\ge0, \quad \forall i\ge1\,.
 \end{equation} 
 This assumption can be justified by the structure of the turning operator. Indeed, to invert $\mathcal{L}$ when solving for the higher-order corrections $f_i$, it is required that $f_i$ belong to the orthogonal complement of the equilibrium space, i.e.\ $f_i \in \langle T_0[\Chem](\hv)\rangle^\perp$. However, for $k \ge 1$, the functions $f_k$ are determined only up to the addition of a component that is constant in $\hv$. The condition \eqref{ass_rhoi} amounts to fixing this indeterminacy by requiring that such constant components vanish, so that no additional mass is carried by higher-order terms.

 If this assumption is not imposed, then at each order in $\varepsilon$ one obtains an additional macroscopic (parabolic) equation governing the $\hv$-independent component of $f_k$ \cite{othmer2002diffusion}. 
 More generally, when mass is not conserved, we still observe that the first-order correction to the density vanishes (cf.\ \eqref{rho1_zero}). Therefore, it is consistent to approximate $\rho$ by $\rho_0$, which satisfies~\eqref{macro_diff_gen} up to an error of order $\mathcal{O}(\varepsilon^2)$.

\end{oss*}

\section{Numerical Simulations}
\label{app:numerical simulations}
Here, we present the additional details to reproduce the numerical results presented in~\Cref{sec:example}. 

While the model is defined on the positive real line ($\xi>0$), numerical simulations are performed on a finite domain $\xi\in[0,L]$ and apply no-flux boundary conditions at both boundaries for the cell density and the chemical species:
\begin{equation}\label{app:noflux_finite}
\partial_\xi\Chem=\partial_\xi\mathcal{B}=0, \quad \xi=0,L.
\end{equation}
The size is taken to be large, {\it i.e.,} $L=400\gg1$, not to influence the invasion dynamics. The values of the parameters used in the simulations are given in Table~\ref{tab:model parameters}.

We solve~\eqref{app:1D_model}-\eqref{app:initial conditions} numerically using the method of lines. Specifically, we first discretise~\eqref{app:chemoattractant dynamics}-\eqref{app:dynamics cell density} on the finite spatial domain $\xi\in[0,L]$, using finite volumes. To this end, we rewrite~\eqref{app:1D_model} in the general form
\begin{equation}
    \frac{1}{\tau_\varphi}\frac{\partial\varphi}{\partial \tau}=-\partial_\xi(\varphi F_\varphi[\boldsymbol{\varphi}])+R_\varphi[\boldsymbol{\varphi}],
\end{equation}
where $\varphi$ indicates a general dependent model variable, while the vector notation $\boldsymbol{\varphi}=[\Chem,\mathcal{B},\rho]$ indicates dependency of functions on (possibly) all tree variables. While the operator $F_\varphi$ describes the speed for variable $\varphi$, the function $R_\varphi$ captures the corresponding reaction term. In this formulation standard diffusion of $\Chem$ and $\mathcal{B}$ is captured by defining $$F_{\varphi}=-\partial_\xi\log\varphi,\quad \varphi\in\left\{\Chem,\mathcal{B}\right\}.$$ 

We divide the domain into $H=1000$ cells, $[\xi_{j}, \xi_{j+1}]$, of equal width $h := \xi_{j+1} - \xi_{j}$, and with centres $\xi_{j+1/2}= (j+1/2)h$ for $j=0,1,\ldots,H-1$. We set $\xi_0=0$ and $\xi_H=L$. Following~\cite{watts2026coupled}, we approximate the value of the solution at each cell centre $\xi_{j+1/2}$ as the cell average:
\begin{equation}\label{eq:general Finite volume}
\varphi(\xi_{j+1/2}, t) \approx \bar{\varphi}_j(t) := \frac{1}{h} \int_{\xi_{j}}^{\xi_{j+1}} \varphi(\xi, t) \, \d\xi, \quad j=0,\ldots, H-1, \quad \varphi\in\left\{\rho,\Chem,\mathcal{B}\right\}.
\end{equation}
We integrate~\eqref{eq:general Finite volume} over each cell mesh, and approximate the nonlinear reaction to obtain an approximate evolution equation for the dynamics of $\bar\varphi_j$:
\begin{equation}\label{eq:discretised time evolution}
    \frac{1}{\tau_\varphi}\frac{\partial\bar{\varphi}_{j}}{\partial \tau} \approx -\left[\varphi F_\varphi[\boldsymbol{\varphi}]\right]^{\xi_{j+1}}_{\xi_j}+R_\varphi[\bar{\boldsymbol{\varphi}}_j], \quad j=0,\ldots, H-1.
\end{equation}
To approximate the cell fluxes at the edge of the cell $J_j(t)=\varphi(\xi_j,t) F[\boldsymbol{\varphi}](\xi_j,t)$, we use an upwind scheme; namely:
\begin{equation}
J_j=\bar{\varphi}_{j-1}(F_\varphi^j[\bar{\boldsymbol\varphi}])_++\bar{\varphi}_{j}(F_\varphi^j[\bar{\boldsymbol\varphi}])_-,
\end{equation}
where $(\cdot)_+$ and $(\cdot)_-$ indicate the positive and negative part operators, respectively. Finally, we approximate the speed at cell edges via the no-flux conditions and using a central finite difference approximation of first-order derivatives. Specifically,
\begin{equation}
    F_\varphi^j=\begin{cases}
    0, &\quad j=0,H\\
    -\dfrac{\log\bar{\varphi}_{j}-\log\bar{\varphi}_{j-1}}{h}, &\quad j=1,\ldots,H-1
    \end{cases}, \quad \varphi\in\left\{\Chem,\mathcal{B}\right\},
\end{equation}
for the chemoattractant and its source, and
\begin{equation}\label{app: discretised flux cell density}
    F_\rho^j=\begin{cases}
    0, &\quad j=0,H\\
    U_j+\varepsilon\tilde{U}_j, &\quad j=1,\ldots,H-1
    \end{cases}, 
\end{equation}
for the cell density. In~\eqref{app: discretised flux cell density}, we approximate the cell speed as
\begin{align}
    U_j&= V_T\left(\Delta_h \bar{S}_j\right)-\nu\Lambda \frac{\bar{\varphi}_{j}-\bar{\varphi}_{j-1}}{h},\\[0.2cm]
    \tilde{U}_j&=-D_T\left(\Delta_h \bar{S}_j\right)\dfrac{\log\bar{\varphi}_{j}-\log\bar{\varphi}_{j-1}}{h}+V'_T\left(\Delta_h \bar{S}_j\right)\frac{S'(\bar{\Chem}_j)\partial_\tau \bar{\Chem}_j-S'(\bar{\Chem}_{j-1})\partial_\tau\bar{\Chem}_{j-1}}{h}\\[0.2cm]
    &\quad-D'_T(\Delta_h \bar{S}_j) \Delta_h^2\bar{S}_j+U_jV'_T(\Delta_h \bar{S}_j) \Delta_h^2\bar{S}_j,\\[-0.4cm]\notag
\end{align}
where $\partial_\tau \Chem_j$ is defined by~\eqref{eq:discretised time evolution}, $\Delta_h \bar{S}_j=h^{-1}(\bar{S}_j-\bar{S}_{j-1})$, and $\Delta^2_h\bar{S}_j= h^{-2}(\bar{S}_j-2\bar{S}_{j-1}+\bar{S}_{j-2})$ is the discretised Laplacian, where the values of $\bar{S}_{-1}:=\bar{S}_0$ and $\bar{S}_H:=\bar{S}_{H-1}$ to account for the no-flux boundary condition for the chemoattractant~\eqref{app:noflux_finite}. 

We integrate the discretised set of ODEs~\eqref{eq:discretised time evolution} using the \emph{Tsitouras 5/4 method}, implemented in the \texttt{DifferentialEquations.jl} package in Julia~\cite{rackauckas2017differentialequations}.
   
%\paragraph{References}
\printbibliography

\end{document}